\documentclass{agujournal2018}
\journalname{JAMES}
\usepackage[]{graphicx}
\usepackage{overpic}
\usepackage{setspace}
\usepackage{color}

\begin{document}

\title{A statistical model for isolated convective precipitation events}

 \authors{Christopher Moseley\affil{1},
 Olga Henneberg\affil{2}, Jan O. Haerter\affil{2}}

\affiliation{1}{Max Planck Institute for Meteorology, Hamburg, Germany}
\affiliation{2}{Niels Bohr Institute, Copenhagen University, Blegdamsvej 17, 2100 Copenhagen, Denmark}

\correspondingauthor{Christopher Moseley}{christopher.moseley@mpimet.mpg.de}

\begin{keypoints}
\item Large eddy simulations
\item Convective precipitation
\item Storm tracking
\end{keypoints}


\begin{abstract}
To study the diurnal evolution of the convective cloud field, we develop a precipitation cell tracking algorithm which records the merging and fragmentation of convective cells during their life cycles, and apply it on large eddy simulation (LES) data. 
Conditioning on the area covered by each cell, our algorithm is capable of analyzing an arbitrary number of auxiliary fields, such as the anomalies of temperature and moisture, convective available potential energy (CAPE) and convective inhibition (CIN). 
For tracks that do not merge or split (termed {\it solitary}), many of these quantities show generic, often nearly linear relations that hardly depend on the forcing conditions of the simulations, such as surface temperature. 
This finding allows us to propose a highly idealized model of rain events, where the surface precipitation area is circular and a cell's precipitation intensity falls off linearly with the distance from the respective cell center. 
The drop-off gradient is nearly independent of track duration and cell size, which allows for a generic description of such solitary tracks, with the only remaining parameter the peak intensity.
In contrast to the simple and robust behavior of solitary tracks, tracks that result from merging of two or more cells show a much more complicated behavior. 
The most intense, long lasting and largest tracks indeed stem from multi-mergers --- tracks involved in repeated merging. 
Another interesting finding is that the precipitation intensity of tracks does not strongly depend on the absolute amount of local initial CAPE, which is only partially consumed by most rain events.
Rather, our results speak to boundary layer cooling, induced by rain re-evaporation, as the cause for CAPE reduction, CIN increase and shutdown of precipitation cells.
\end{abstract}

\section{Introduction}\label{sec:intro}
Recent studies, using both observational and modeled data, argued that future warmer climate conditions may result in an intensification of convective precipitation \citep{OGorman:2009,Lenderink:2008,westra2014future,lenderink:2017}, potentially increasing the risk of flood \citep{kendon2014heavier}. 
Historical high-resolution data show that convective precipitation intensities are particularly sensitive to temperature changes \citep{berg2013strong,Lenderink:2008,Lenderink:2009,molnar2015storm}, but the exact mechanisms causing extreme convective precipitation or its temperature dependence are, to date, not fully understood.

A recent paper suggested, that disappearance of local CIN in a given location is a necessary prerequisite for the onset of convection there \citep{moseley2016}.
Negative buoyancy contributions, defined as CIN, are hence a plausible indicator of times and locations where convection is suppressed. 
We therefore work with local definitions of both CIN and CAPE.
CAPE is a traditional predictor for convective 
intensity and updraft speed 
, repeatedly used in subgrid closure schemes for convective parametrization in large scale models \citep{arakawa1974interaction,arakawa2004cumulus}. 
Failure of convective parametrizations was found to be more likely under non-equilibrium conditions when CAPE is rapidly consumed with the onset of convection and not balanced by the generation by large scale processes \citep{Done2006, Zimmer2011}. 
The exact predictive meaning of CAPE is hence far less clear than that of CIN, an issue we address in this work.
Indeed, CAPE suffers from being 
a conceptual, idealized 
quantity, based on the adiabatic ascent of a test parcel originating at a certain height above the surface.
CAPE therefore neglects the effects of mixing, e.g. convective entrainment during ascent, by which a test parcel 
loses some of its buoyancy and potential energy is converted into vertical motion with an efficiency reduced by almost 50\,\% compared to an undiluted parcel \citep{Zipser2003}. Additionally, less of the water vapor carried by the parcel 
condenses. 
The efficiency of CAPE is furthermore dependent on its vertical distribution as parcels are accelerated faster when CAPE is distributed over a shallower layer \citep{Blanchard1998}. \\
Here we aim to address the interplay between CAPE, CIN and local precipitation production as well as other thermodynamic quantities. 
As mentioned, we therefore define CAPE and CIN as local quantities, computed separately for every model column. 
We deliberately focus on CAPE and CIN because they are pure thermodynamic quantities. 
The local thermodynamic conditions before and during the initiation of precipitation cells are mainly influenced by surface latent and sensible heat fluxes as well as the redistribution of moisture and temperature within the boundary layer. 
By these processes, certain locations may benefit from increased near-surface buoyancy, thus setting off an updraft there. 
With the onset of precipitation the reduction of CAPE can be related to the precipitation rate associated with a latent heat release that eliminates the temperature differences between the parcel and the ambient air temperature \citep{Done2006}. 

Storm tracking algorithms have proven useful in studying the interaction of convective rain cells \citep{dawe2013direct}.
In recent years, a number of methods for the tracking of individual clouds and convective storms have been developed. 
These methods are able to identify single convective cells and follow their evolution throughout the life cycle, that is, from their formation to their dissolution. 
All of these methods have been developed for different purposes and are therefore specialized in one way or the other.
Elaborate and optimized storm tracking methods in two and three dimensions have been developed for the purpose of nowcasting thunderstorms \citep{Dixon:1993,Hering:2005,kober2009};
other methods are rather designed for the tracking of clouds and thermals, to study the cloud statistics in shallow convection \citep{Heus:2013,Heilblum:2016}, or in deep convection \citep{tsai:2017,senf:2017}.

Our current tracking method focuses on the life cycle of convective precipitation tracks in both observed and simulated 2D data, such as 2D radar precipitation observation products and surface rainfall from large eddy simulation (LES) output.
We build on the {\it Iterative Rain Cell Tracking (IRT)} originally developed for the analysis of radar data to study the scaling of the intensities of single rain cells with near-surface temperature \citep{moseley2014}. 
Compared to other methods cited above, IRT is simple, but it is able to distinguish merging and fragmentation incidents and therefore caters to statistical analysis of different track types.

Mergers have been described before, both observationally \citep{byers1949thunderstorm,simpson1980cumulus} and by computer simulations \citep{tao1989further,glenn2017connections}, suggesting that combined cells can produce more intense precipitation, and that merging can lead to larger detrainment heights \citep{glenn2017connections}.
In IRT, instantaneous contiguous objects of precipitation are identified and checked for overlaps with the respective consecutive time steps \citep{moseley2014}.
IRT capitalizes on the fact that, even under large scale advection, larger objects mostly overlap from one time step to the next, thereby allowing to identify tracks formed by the objects with overlap.
By iterating, also smaller rain cells, which often do not overlap when they are advected, are captured. 


An approach similar to IRT has more recently been applied to the analysis of radar data over the Netherlands -- finding that the temperature scaling of cell intensities depends on the cell size \citep{Lochbihler:2017}.
IRT has further been applied to precipitating convective updrafts generated by idealized large eddy simulations (LES) \citep{moseley2014}. 
The main finding there was that tracks resulting from the merging of previous tracks react much more strongly to forcing conditions when compared to tracks that did not interact, so-called {\it solitary tracks}. 
The reason for this insensitivity of the solitary tracks to surface forcing could not yet clearly be identified, but it could be speculated that they are mainly driven by the feedback between the boundary layer and the free atmosphere, which is largely independent of the boundary layer height.

To address the relation between CAPE, CIN and precipitation, IRT is here extended to record also an arbitrary number of auxiliary fields by conditioning on surface precipitation intensity (Sec.~\ref{sec:methods}).
We hence yield time evolution both for the main tracking field and all auxiliary fields.
In Sec.~\ref{sec:results} we describe our results, including overall track statistics and a characterization of the track life cycle.
We then discuss the relation between CAPE, CIN and near-surface temperature changes, which leads us to propose a simple statistical model, that can capture the relation between cell duration, cell maximum intensity and cell area.
We finally discuss the implications and possible extensions and conclude (Sec.~\ref{sec:discussion}).


\section{Tracking method and model simulation data}\label{sec:methods}

Here we define our tracking method including the basic procedure of identifying precipitation {\it objects} and the associated fields (Sec. \ref{subsec:methods:tracking}). 
The tracking is applied to high-resolution data produced from a typical LES model described in Sec.~\ref{subsec:methods:model}.

\subsection{Tracking method}\label{subsec:methods:tracking}

\noindent
{\bf Objects and tracks.} The algorithm diagnoses precipitation cells, or contiguous areas defined by any other field, as disparate entities in space, in the following called {\it objects}. Precipitating and non-precipitating areas are separated by a fixed threshold $I_{min}$. Throughout this paper, we choose a threshold of $I_{min}=1\;mm\,h^{-1}$ since this value corresponds to the typical limit of detectability in radar measurements \citep{moseley2014}. 
After object identification, the tracking algorithm links objects between two consecutive output time steps when they overlap.
Such overlapping objects are then considered part of the same {\it track}. By definition, we require an object to consist of at least four grid boxes, and a track must be at least two time steps long (i.e. tracks that are only one time step long are neglected). We define the track {\it lifetime} as the number of time steps multiplied by the output interval of the input fields, e.g. to a track that is 6 time steps long we would assign a lifetime of 30 minutes, since our LES data have output interval of 5 minutes. Note that the term "object" is used here for a precipitation cell at a given instant in time, while a "track" is an entity with a given lifetime, i.e. it links all objects at different time steps together that belong to the same precipitation event.

\noindent
{\bf Iteration.} Under rapid advection, smaller objects often do not overlap from one time step to the next although they might belong to the same updraft process. 
To remedy this shortcoming, an iterative procedure is applied: 
Object identification is performed once, and a mean advection velocity field of the moving objects is diagnosed from the tracking result. Subsequently, the tracking is repeated by taking into account the diagnosed velocity field, such that each object is displaced by $\Delta {\boldmath{r}}\equiv \Delta t \cdot {\boldmath{v}}$, where $\Delta {\boldmath r}$ is the displacement, $\Delta t$ the data output time step, and $\boldmath{v}$ is the velocity. 
This results in an improved match with the corresponding object of the consecutive time step (for details, see \cite{moseley2014}).
Usually, the iterative procedure has to be repeated several times until the diagnosed velocity converges. 

\noindent
{\bf Merging and fragmentation.} The main challenge for the tracking algorithm is the handling of merging and fragmentation incidents. 
When very small objects combine with much larger ones, one might not consider the resulting track distinct from the larger of the two.
We therefore introduce a parameter $\theta$, termed the {\it termination sensitivity}, that can be used to distinguish if a merging/fragmentation incident entails the termination of all involved tracks, or if the largest track is continued. 
Specifically, we define as follows:
\begin{itemize}
\item{\it Merging incident:} two or more objects at timestep $t$ overlap with one object at time step $t+1$. 
The algorithm determines the areas $A_i$ of the largest object $O_i$ and $A_j$ of the second largest object $O_j$ at time $t$. 
If $A_j/A_i<\theta$, then $O_i$ is continued as the merged object at time $t+1$, while $O_j$ and (if present) all other smaller objects are terminated. 
Otherwise, if $A_j/A_i\geq \theta$, all objects at time $t$ are terminated, and the merged object at time $t+1$ is initiated as a new track and labeled as a track initiated from merging. 
\item{\it Fragmentation incident:} The definition is analogous, but for the case where parts of one track separate from an existing track. 
Now the areas of the fragments are compared, again using the comparison of the largest and second to largest area, as for mergers from one time step to the next. In principle, a fragmentation incident is the time reverse of a merging incident.
\end{itemize}

the parameter $\theta$ takes values $0\leq\theta\leq 1$. 
If $\theta=0$, every merging and fragmentation incident leads to the termination of all involved tracks, and to a new initiation of all resulting tracks. The main objective for the introduction if $\theta$ is the reduction of noise: If a very small object splits off or merges into a much larger one, it can be avoided that the large track is immediately terminated by such an event by choosing a non-zero value for $\theta$. For $\theta=1$, upon merging or fragmentation, at least the track with the largest object area at the time of the merging continues.

\noindent
{\bf Book keeping.}
Tracks are labeled by the type of their initiation and termination. We use the notation {\it X-Y}, where {\it X}$\in$\{s,m,f,a\} denotes the type of initiation, i.e. as a new {\it solitary} event (s), as a result of a {\it merging} event (m), or as a {\it fragment} of a splitting-up of another track (f), and {\it Y}$\in$\{s,m,f,a\} denotes the type of termination, i.e. dissolution as a {\it solitary} event (s), by {\it merging} with another track (m), or by {\it fragmentation} or breaking-up into other tracks (f). For instance, {\it s-s} means all tracks that begin and end as solitary and thus do not interact with other tracks, while {\it m-s} denotes tracks that begin as a merging result, but terminate by dissolution. The symbol {\it a} is used as a place holder for all of the three types s,m,f for either initiation or termination, e.g. {\it s-a} denote tracks that begin as solitary but end in any of the three possibilities. In the following we will refer to tracks of type {\it s-s} simply as "solitary", as they will constitute the main part of  discussion.
Note that the terms {\it initiation} and {\it termination} are used here only in association with the tracks as mathematical objects identified by the IRT algorithm, and not with the physical rain events as such, which of course do not terminate when merging and fragmentation incidents happens. The number of tracks that are detected for each track type depends on the choice of the parameter $\theta$, as will be discussed in Section \ref{sec:categorization}.

\noindent
{\bf Auxiliary variables.}
Area mean, maximum and minimum of any additional fields are recorded for the areas defined by the main tracking variable (here: surface precipitation). 
Here we record the following additional variables: 
Anomalies, that is, subtracting the current domain mean, of temperature in the lowest model level, convective available potential energy (CAPE) and convective inhibition (CIN).
CAPE and CIN are defined at a gridbox level using an adiabatically lifted test parcel from the surface to the level of neutral buoyancy.

\noindent
{\bf Boundary conditions.}
IRT can be applied on data both with periodic (as is the case here) or open boundary conditions (such as for remote sensing data or limited area simulations). 
Further, IRT can also handle missing values which may occur in observational data. 

\noindent
{\bf Fortran 90 source code.}
We make the IRT program code publicly available in combination with this paper. The source code, including a user's manual and a tutorial, can be downloaded via the URL\\
https://github.com/christophermoseley/iterative\_raincell\_tracking

\subsection{Model simulation data} \label{subsec:methods:model}

We simulate an idealized convective diurnal cycle using the University of California, Los Angeles (UCLA) Large Eddy Simulation (LES) model \citep{stevens2005}. 
The domain size is 1024$\times$1024 grid boxes with a horizontal grid spacing of 200 $m$, with 75 vertical levels which stretch from a spacing of 100 $m$ near the surface to 400 $m$ at the model top, located at 16.5 $km$. 
The simulation is initialized by horizontally homogeneous temperature and moisture profiles. 
The temperature profile starts with 21 $^\circ C$ at the lowest model level with a lapse rate of $6.6\;K km^{-1}$ below $11\;km$ and $3\;K km^{-1}$ above.
The profile of relative humidity starts with 65\% in the first model level, linearly increases by 12\% $km^{-1}$ below $2\;km$, decreases by 12.5\% between 2 and $4\,km$, by 2\% $km^{-1}$ between 4 and 10 $km$ and by 12\% $km^{-1}$ higher up.

The diurnal cycle is imposed by a varying surface temperature ($T_{surf}$) profile following
$T_{\rm surf}(t)=T_0+\Delta T\sin((t-6)\pi/12)$, where $t$ denotes the time in units of hours after midnight, and the solar insolation at a latitude of 52$^\circ$ N. 
$T_0$ is the daily average surface temperature, and was varied between $T_0=23,25$ and $27$ $^\circ C$, denoted in the following as the {\it CTR}, {\it P2K} and {\it P4K} simulations.
The temperature amplitude $\Delta T=10\;K$.
An additional run, which includes large-scale advective forcing as well as a simulated vertical lifting is denoted as {\it OMEGA}. For all simulations, we chose an output interval of 5 minutes for the full 3D prognostic fields.
A more detailed description of the simulation and model setup is given in \cite{moseley2016}, where it has been shown that in the simulations with increased surface temperature, the convective life cycle starts earlier, covers a larger part of the domain, and includes more merging incidents between the cells: While convective cells in CTR are mostly isolated, they interact more strongly in the P2K and even stronger in the P4K simulations.

We include the simulation named OMEGA into our analysis as it includes homogeneous large scale wind shear and is therefore qualitatively different than the simulations without shear: Precipitation objects are elongated in the flow direction and grow larger during the course of the day. The iteration process mentioned in Sec. \ref{subsec:methods:tracking} is applied on the OMEGA simulation only, but not on CTR, P2K, and P4K, since there is no background flow in the latter these simulations. Although the focus of the following results section is on the three simulations without large scale forcing, we include OMGEA for the sake of completeness, and argue that the behavior of convective events is fundamentally different in the case of wind shear that would require a special consideration.

\subsection{Calculation of local CAPE and CIN} \label{subsec:methods:cape}

For the calculation of CAPE and CIN, in every model column we lift an imaginary air parcel along a pseudoadiabat, starting from the lowest model level, up to the model top. CAPE is then given by
\begin{equation}
\label{eq:cape}
CAPE = \int g\left( \frac{T_{v,parcel}-T_{v}}{T_v}\right) dz \; ,
\end{equation}
where $T_v$ is the virtual temperature of the air in the column, $T_{v,parcel}$ is the virtual temperature of the parcel, and the integral is taken over all values of $z$ above condensation level where $T_{v,parcel}>T_v$, i.e. where the parcel has positive buoyancy. CIN is given by
\begin{equation}
\label{eq:cin}
CIN = \int g\left( \frac{T_v-T_{v,parcel}}{T_v}\right) dz \; ,
\end{equation}
where the integral is taken over all values of $z$ below the level of free convection where $T_{v,parcel}<T_v$, i.e. where the parcel has negative buoyancy.

We approximate the pseudoadiabat by a dry adiabat below condensation level, and above condensation level by integration of a temperature lapse rate given by \cite{emanuel}:
\begin{equation}
\label{eq:lapse}
-\frac{dT}{dz} = \frac{ \Gamma_d +
\frac{L_v}{c_p} \frac{R_d/R_c\times e_{sat}pg}{
(p-0.378e_{sat})^2 R_d T} }{
1+\frac{L_v}{c_p}\left( \frac{R_d/R_v}{p-0.378e_{sat}} + \frac{0.378 R_d/R_v\times e_{sat}}{p-0.378e_{sat}}\right)
\frac{de_{sat}}{dT} } \; ,
\end{equation}
where $R_d$ and $R_v$ are the specific gas constants of dry air and water vapor, respectively, $L_v$ is the latent heat of evaporation, $c_p$ is the specific heat capacity of dry air at constant pressure, $e_{sat}$ is the saturation vapor pressure, $p$ is the air pressure, and $\Gamma_d=0.0098Km^{-1}$ is the dry adiabatic lapse rate.

\section{Results}\label{sec:results}

\subsection{Categorization of track types}\label{sec:categorization}

\begin{figure}
    \centering
    \includegraphics[width=14cm,clip]{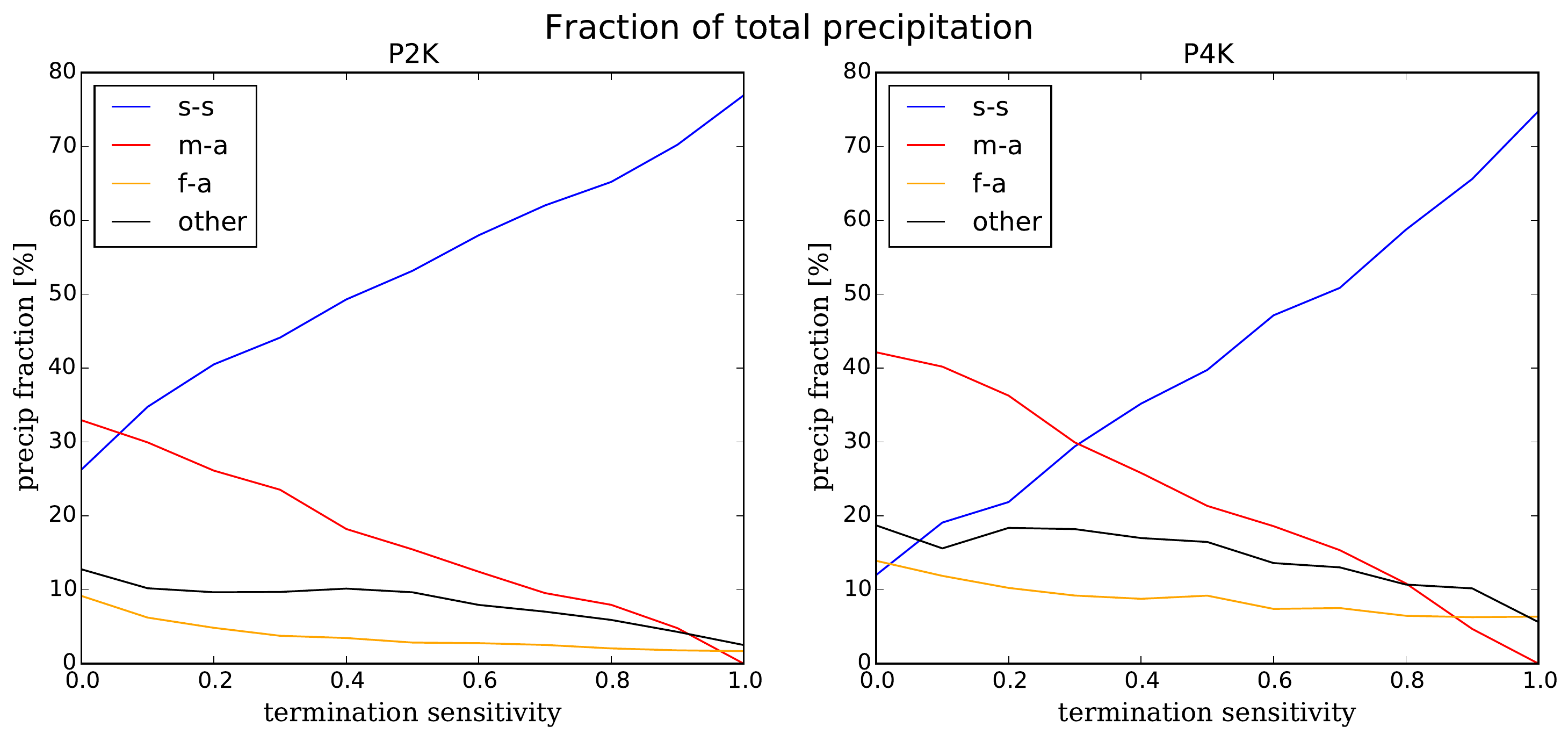}
    \caption{Fraction of total accumulated rainfall for track types {\it solitary-solitary}, {\it merger-all}, {\it fragment-all}, and remaining types ({\it other}), versus the termination sensitivity $\theta$, for the simulations {\it P2K} (left panel) and {\it P4K} (right panel).}
    \label{fig:threshold_ratio}
\end{figure}

As mentioned in Sec. ~\ref{sec:methods}, there are three processes by which tracks can be both initiated and terminated: Solitary (s), merging (m) or fragmentation (f), hence yielding nine possible combinations. The number of tracks that are detected within each type depends on the choice of the parameter $\theta$.
To give a basic assessment of the impact of $\theta$, we compute the total precipitation throughout the diurnal cycle for several track categories (Fig.~\ref{fig:threshold_ratio}), that is, pure solitary ({\it s-s}), all tracks initiated through a merging incident ({\it m-a}), tracks initiated during fragmentation ({\it f-a}), and remaining track types (termed: {\it other}), that is, {\it s-m} or {\it s-f}. 
The sum of all components shown  amounts to approximately 80\% for {\it P2K} and 86\% for {\it P4K}. 
The remaining precipitation with intensities below 1 $mm\,h^{-1}$ bypasses the object identification, or belong to tracks that are only one time step long.
Lower threshold values will naturally increase the records of low-intensity precipitation objects.
As Fig.~\ref{fig:threshold_ratio} shows, the fraction of the different categories varies substantially with $\theta$, with the contribution of solitary tracks increasing with $\theta$ at the expense of mergers. For very low values of $\theta$, even very small tracks can lead to such merging and fragmentation incidents which might perturb a clear signal of the interaction, while for $\theta=1$, mergers completely vanish, as each track that a smaller tracks merges into, is always continued and is considered as solitary (vice versa, in a fragmentation incident, the largest fragment always continues the original track). Therefore, in the case $\theta=1$ the number of detected solitary tracks is maximal (see also the supplementary discussion in section \ref{sec:theta} for further details). 

\begin{figure}
    \centering
    \includegraphics[width=14cm,trim={0 2.8cm 0 0},clip]{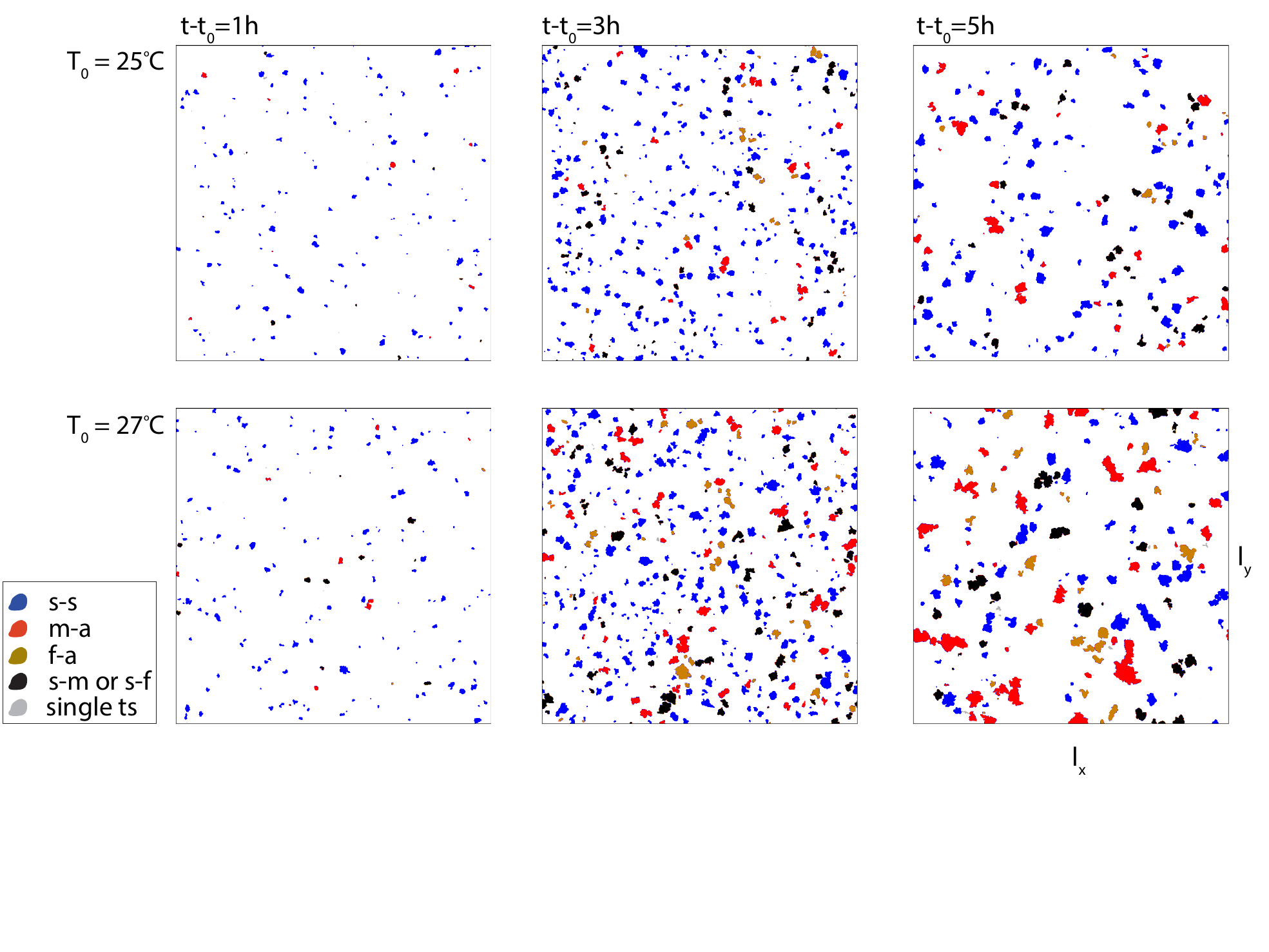}
    \caption{{\bf Time sequences of tracked objects.} Precipitation objects one hour, three hours and five hours after the onset of precipitation for $T_0=25^{\circ}C$ ({\it P2K}) and $T_0=27^{\circ}C$ ({\it P4K}), as labeled, with $\theta=0.5$. 
    Objects are colored by their track type as indicated in the legend. 
    "single ts" labels tracks that lasted for only a single time step.
    $\theta=0.5$, $l_x=l_y=204\;km$.
    }
    \label{fig:trackmask}
\end{figure}

As an example, an instantaneous situation containing objects belonging to all track types mentioned above is shown in Fig. \ref{fig:trackmask}. In this case, we use an intermediate value of $\theta=0.5$, which allows the possibility of mergers.
An example of a simulation with background large scale advection (OMEGA) is shown in Fig.~\ref{fig:trackmask_supp}.
To track rain cells in these data, the iteration feature of the tracking algorithm is applied --- as described in section \ref{subsec:methods:tracking}.
We contrast the advected with the unadvected {\it CTR} simulation at the same surface temperature.

\begin{figure}
    \centering
    \includegraphics[width=6cm,trim={0 2.35cm 0 0},clip]{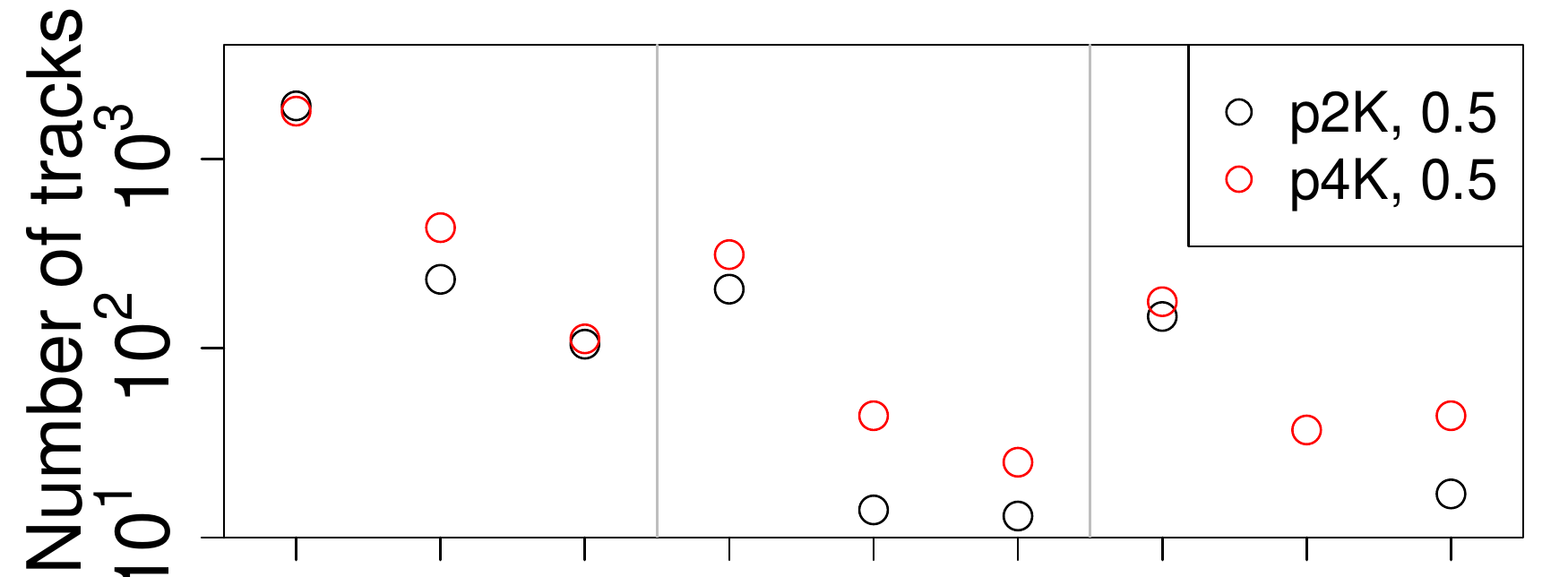}\\
    \includegraphics[width=6cm,trim={0 2.35cm 0 0},clip]{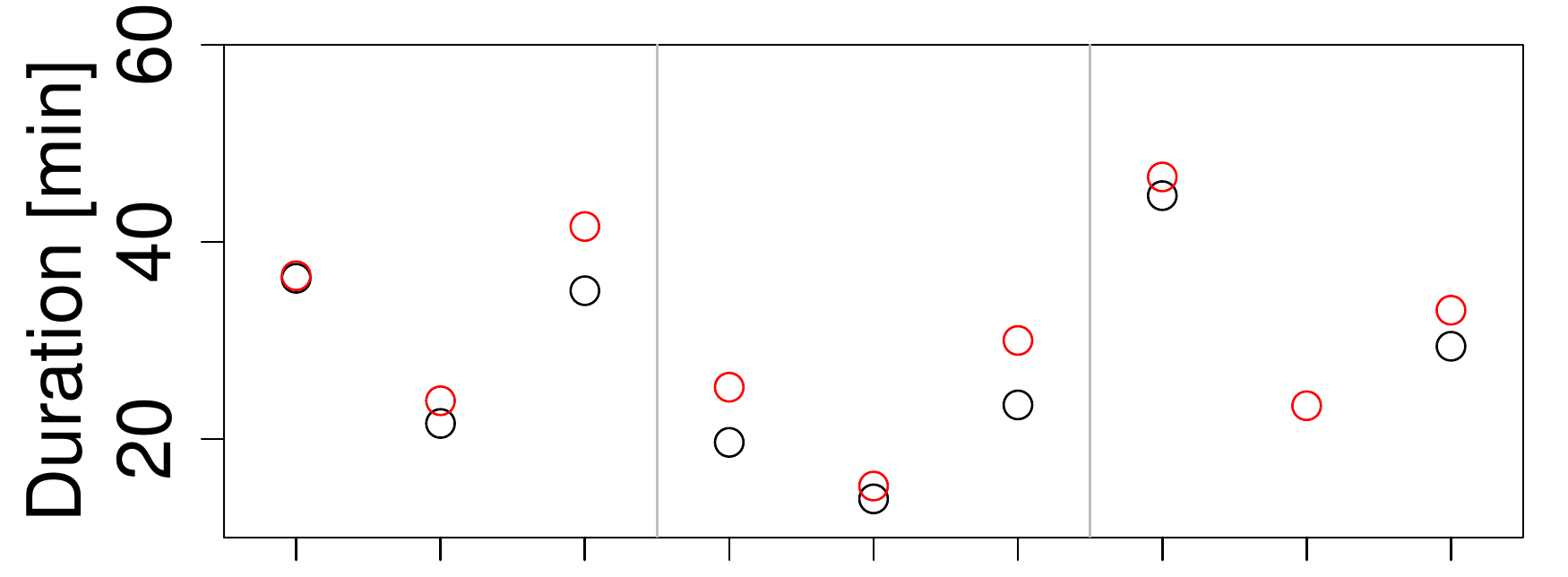}\\
    \includegraphics[width=6cm,trim={0 2.35cm 0 0},clip]{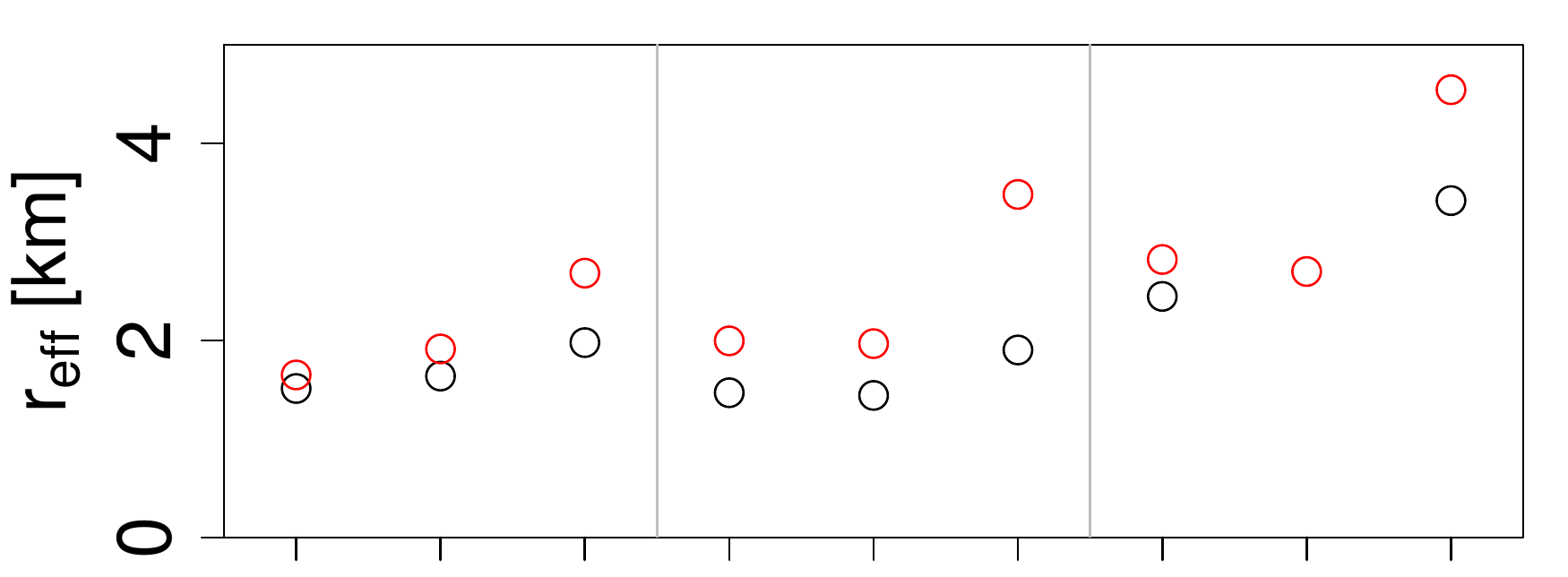}\\
    \includegraphics[width=6cm,trim={0 0cm 0 0},clip]{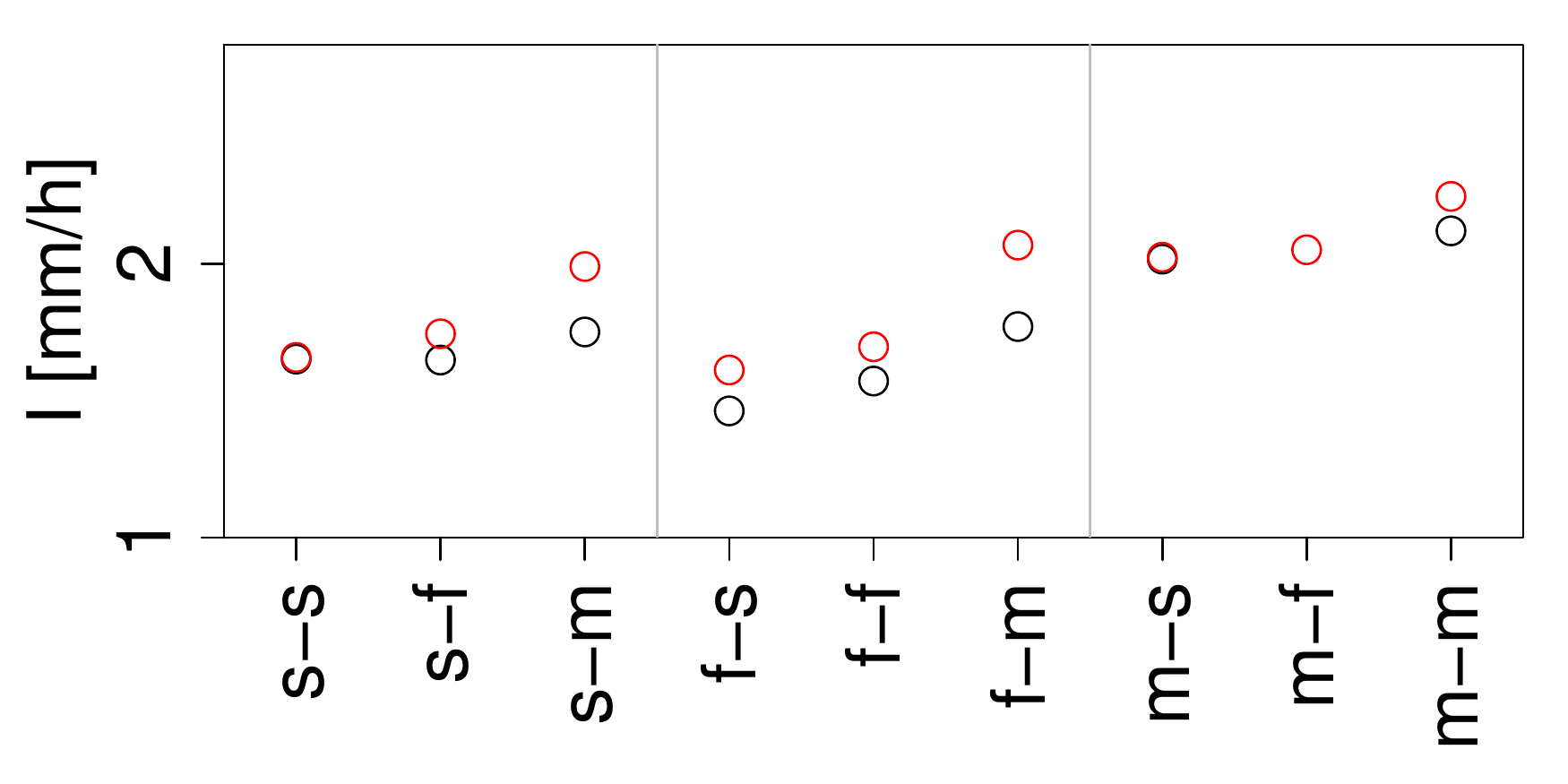}
    \caption{{\bf Relative track properties.}
    Number of tracks, averages of track duration, effective patch radius and precipitation intensity for each track type. 
    Only cases with more than ten data points were considered in the statistics.
    Note the logarithmic vertical axis scale in top panel.}
    \label{fig:track_properties_basic}
\end{figure}

Fig.~\ref{fig:track_properties_basic} shows the total number of detected tracks, and some basic mean statistics, for each track type for the simulations, again for $\theta=0.5$ to include mergers.
Overall, tracks that are initiated as solitary are by far the most abundant.
The second most prominent category are tracks originating as fragments or mergers but ending as solitary.
Tracks both originating and ending as interaction incidents are relatively rare. 
Notably, when comparing the two simulations {\it P2K} and {\it P4K}, which have different surface temperature forcing $T_0$, stronger forcing leads to increase of the number of mergers and fragmentations at the expense of the purely solitary tracks.
The overall number of tracks remains nearly unaffected by the forcing change.

Mean track durations vary between 15 and 45 $min$ and are largest for tracks originating as mergers.
Shortest durations occur for tracks initiated and terminated as fragments --- likely a statistical effect where multiple fragmentation processes take place in quick succession.
As has been shown before \citep{moseley2016}, under the forcing change, solitary tracks do not significantly change their duration, while tracks of all other categories do (mostly increasing for the stronger forcing).

The typical spatial extent of tracks (quantified by $r_{eff}$) is $2$---$4$ $km$ and generally largest for tracks originating from merging. 
Merging not only leads to expectedly larger areas, but even
the area average intensity of mergers increases relative to purely solitary tracks.
Overall largest, and most intense, tracks result from repeated merging ({\it m-m}). 
Further, {\it m-m} tracks increase in number, duration, size and intensity when the forcing is increased.

Together, these findings suggest, that tracks involving some form of interaction (either merging or fragmentation) react to increased forcing.
They do so by increasing duration, spatial extent and precipitation intensity. 
Purely autonomous solitary tracks (i.e. of type {\it s-s}) show essentially no change in any of these properties. 
Furthermore, stronger forcing increases the probability of interference between tracks, a feature reflected by increased track numbers with merging or fragmentation, at the expense of purely solitary tracks. 
This shift, together with the aggravated properties of the interacting tracks, leads to overall more intense and more widespread domain-mean precipitation.

\subsection{Time evolution of convective rain tracks}

\begin{figure}
    \centering
    \includegraphics[height=5.15cm,trim={0 0cm 0 0},clip]{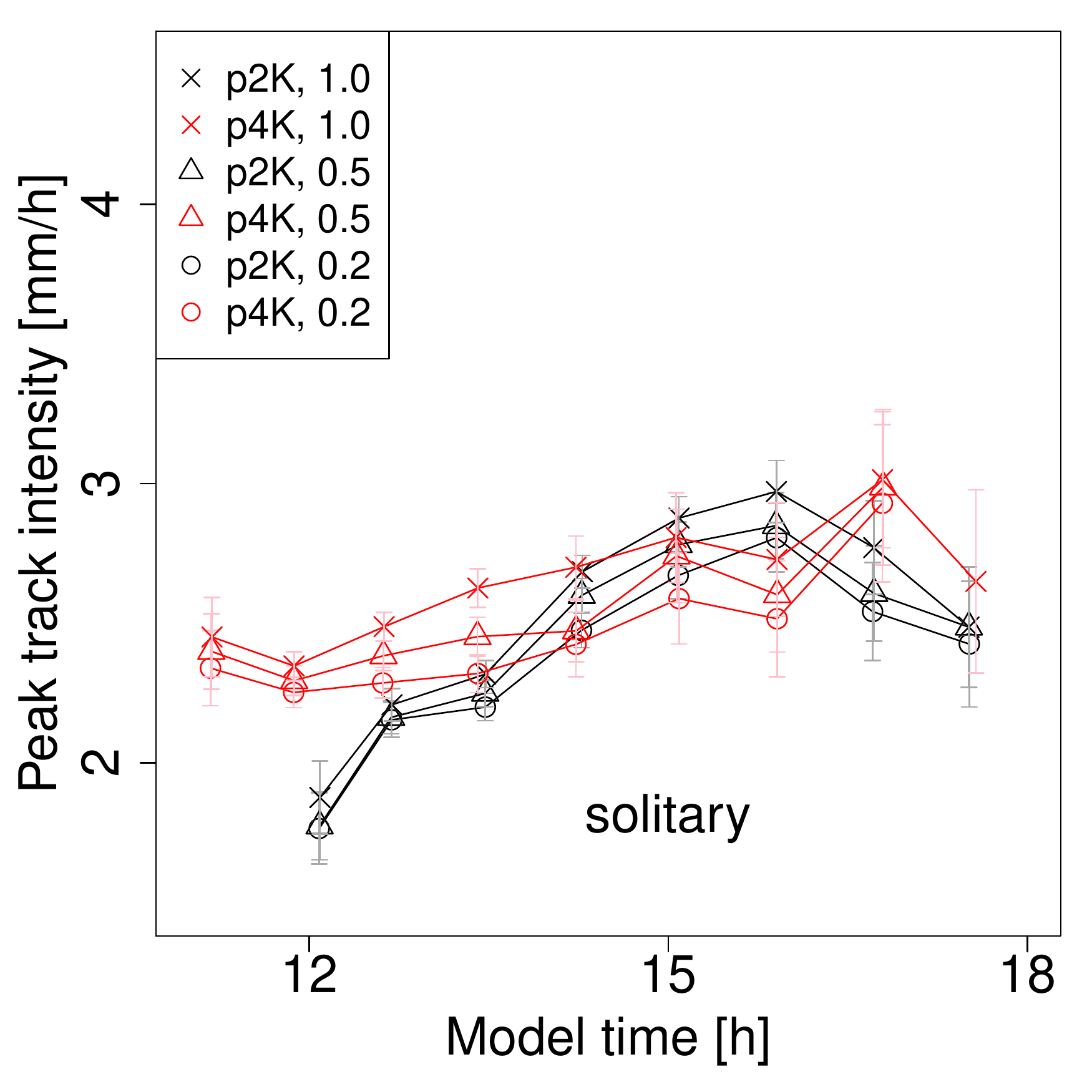}
    \includegraphics[height=5.15cm,trim={2.1cm 0 0 0},clip]{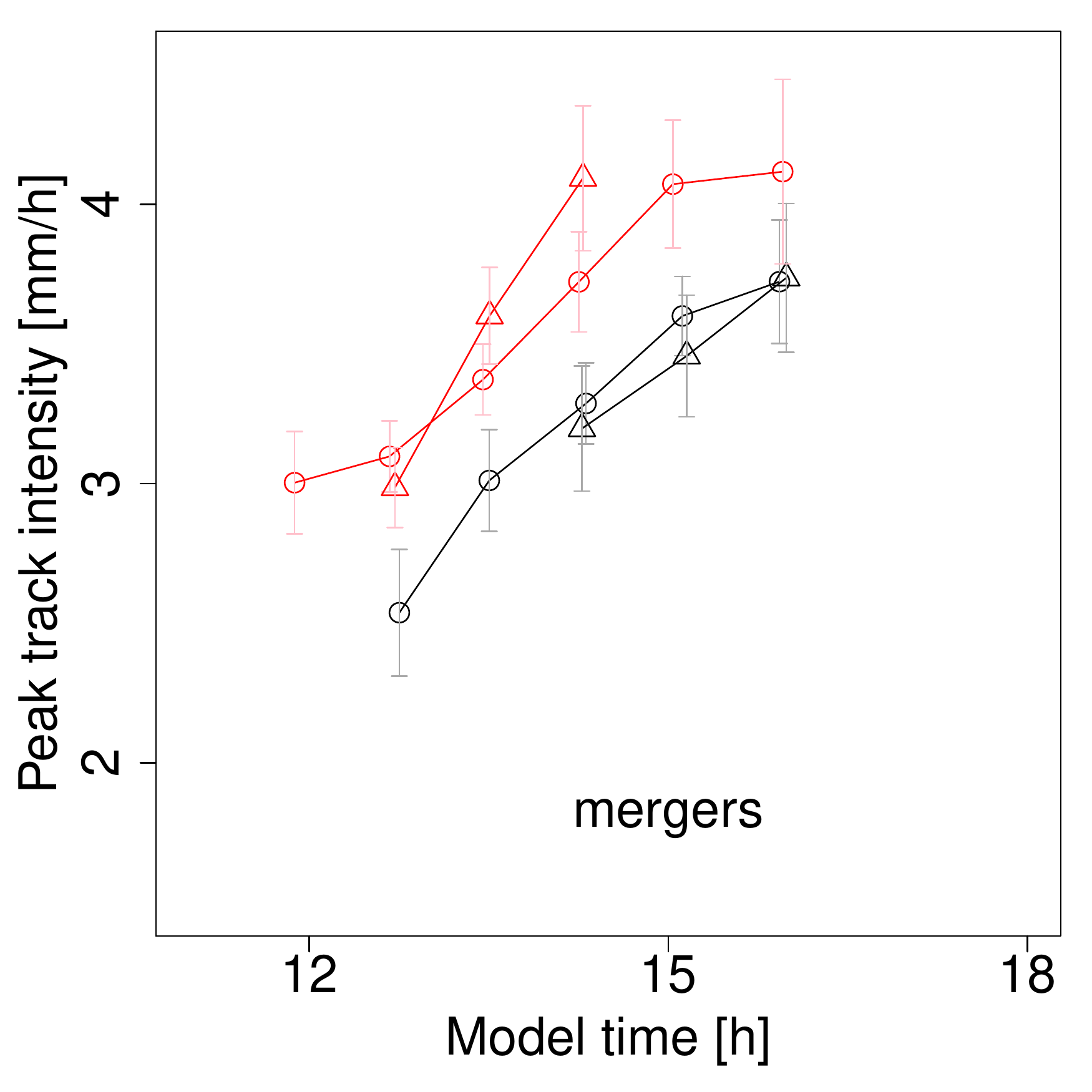}

    \caption{{\bf Peak track intensity vs. time.} Comparison of purely solitary ({\it s-s}, left) vs. mergers ({\it m-a}) tracks (right), for the simulations {\it P2K} and {\it P4K} (colors), and for termination sensitivities $\theta=1.0$, $\theta=0.5$, and $\theta=0.2$ (symbols as shown in legend).
    Note that the case of $\theta=1.0$ is not shown for mergers, since their number vanished in this case.
    The error bars indicate the standard error for each data point. 
    }
    \label{fig:track_duration_vs_time}
\end{figure}

\noindent
{\bf Evolution during the diurnal cycle.}
We find that track intensities steadily increase (with some noise)
during the diurnal cycle (Fig.~\ref{fig:track_duration_vs_time}).
Hence, the longest-duration and most intense tracks are expected in the late afternoon hours. 
The increase in duration is especially pronounced for mergers (type {\it m-a}), which indicates that longer and more intense tracks are formed by merging incidents during the course of the day. 
The intensification of mergers is more pronounced for stronger surface forcing (P4K).  
In addition, results are relatively robust against changes of the termination sensitivity $\theta$ between values of $0.2$ and  $1.0$, especially for solitary tracks. 
However, in the following we are mainly interested in the life cycles of solitary tracks and therefore choose $\theta=1$, as the number of detected solitary tracks becomes largest in this case as stated above.

\begin{figure}
    \centering
    \includegraphics[width=4.5cm,trim={0cm 2.35cm 0 0} ,clip]{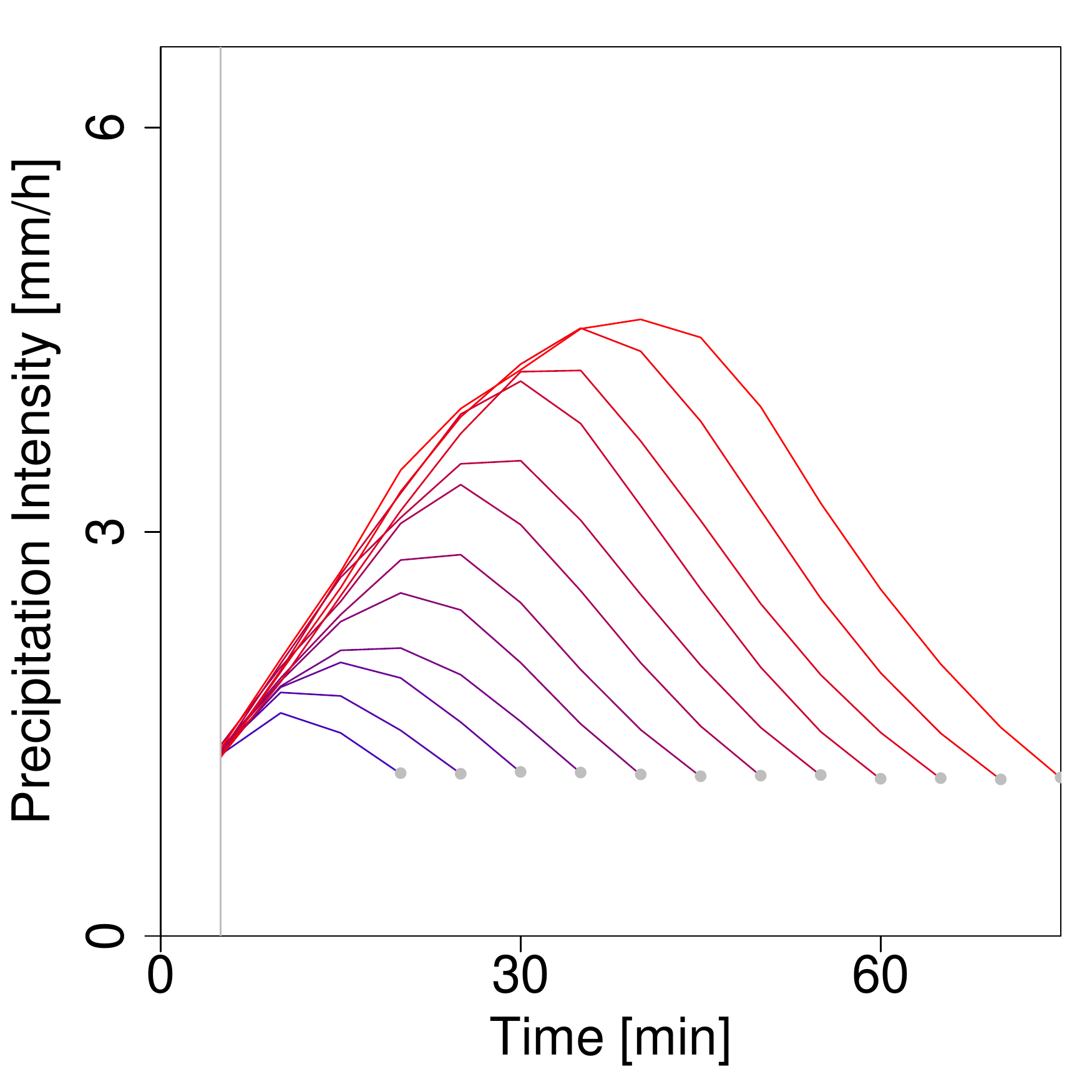}
    \includegraphics[width=4.5cm,trim={0 2.35cm 0 0} ,clip]{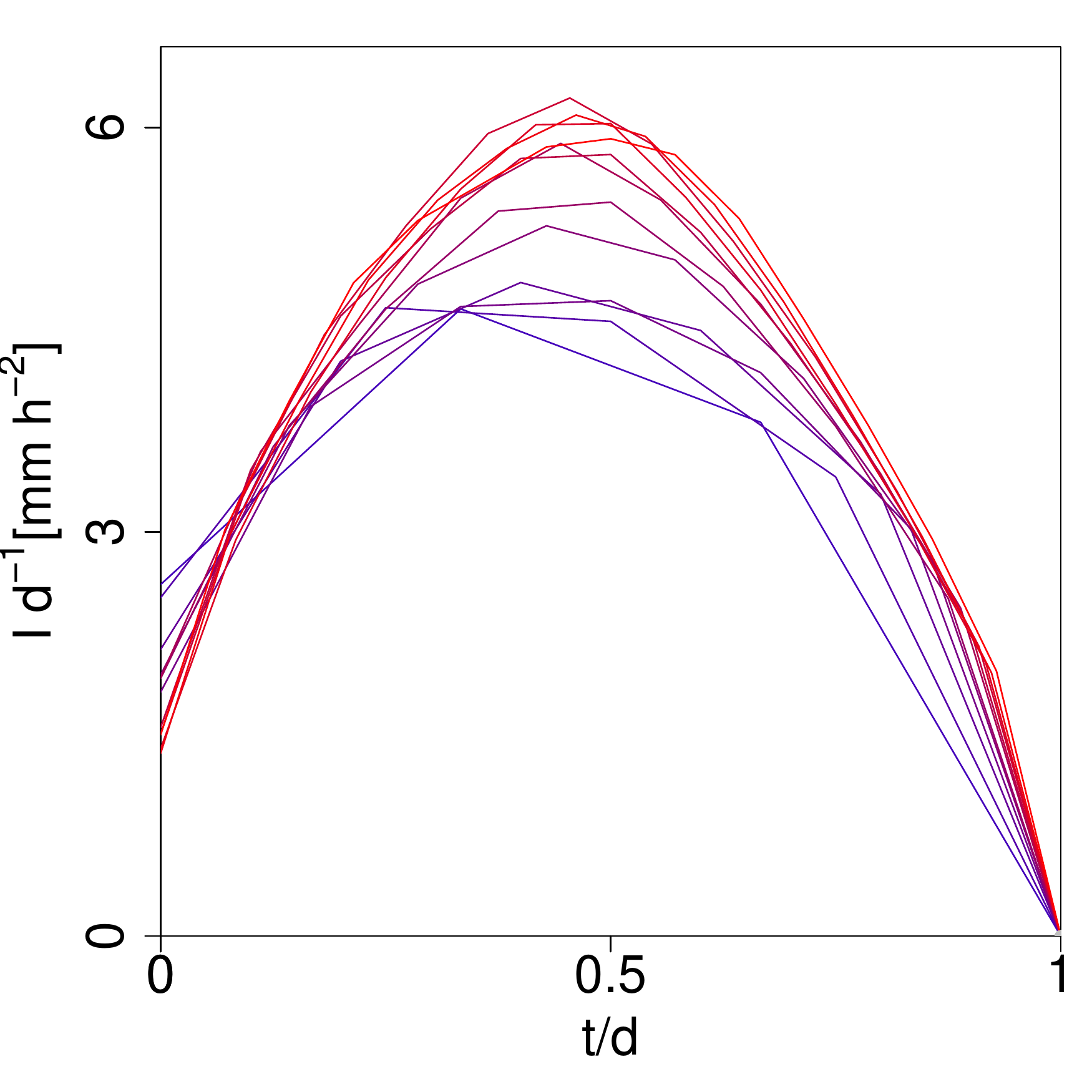}\\
    \vspace{-0.0cm}
    \includegraphics[width=4.5cm,trim={0 0cm 0 0},clip]{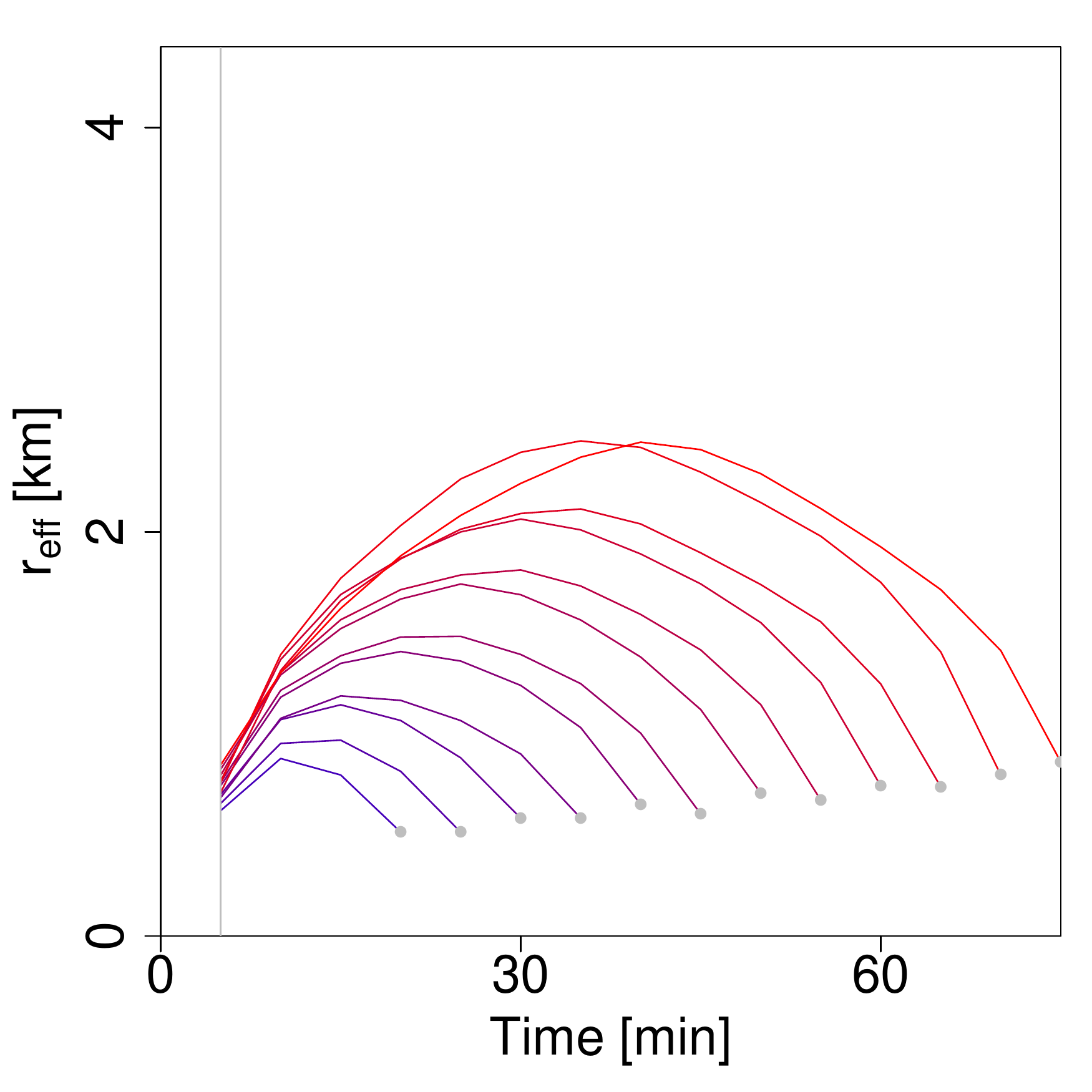}
    \includegraphics[width=4.5cm,trim={0 0cm 0 0} ,clip]{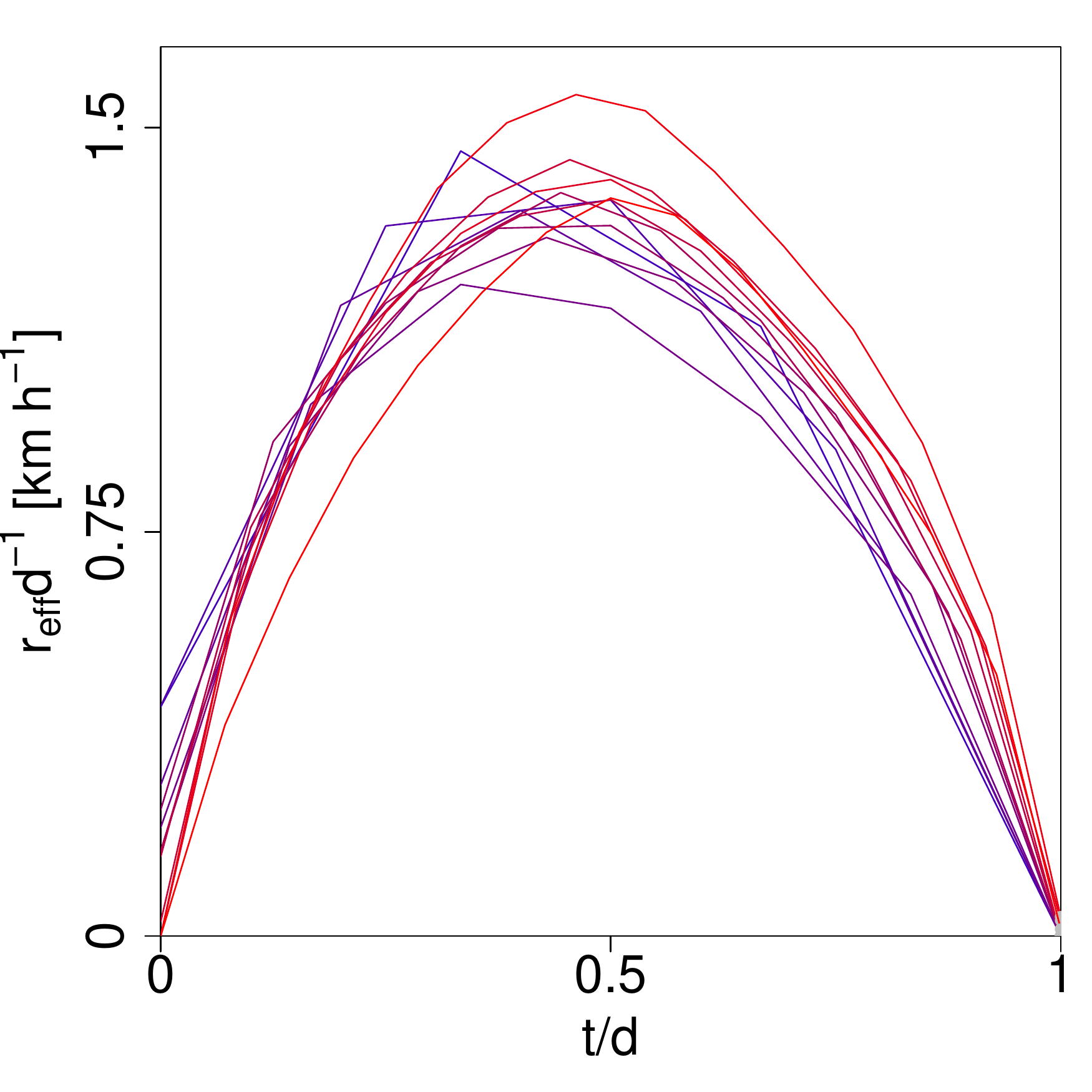}\\
    \caption{{\bf Time dependence across track duration.}
    Precipitation intensity and effective object radius ($r_{eff}\equiv \sqrt{A/\pi}$) vs. time within tracks (left columns), for the P2K simulation and $\theta=1.0$.
    Colors of curves indicate mean track life cycles conditioned on duration, ranging from 10 $min$ (blue) to 90 $min$ (red).
    Right columns: Both axes normalized by total track duration. Intensity $I$ is here defined as precipitation intensity minus the minimal intensity of the corresponding track average (typically the threshold intensity $I_{min}$).
    }
    \label{fig:time_dependence}
\end{figure}

\noindent
{\bf Track life cycles.} 
We now consider how tracks evolve within their life cycles.
Solitary tracks show remarkable systematic properties:
The mean temporal evolution of type {\it s-s} is characterized by a single-peaked structure in precipitation intensity, with the peak occurring approximately after half the track duration (Fig.~\ref{fig:time_dependence}, compare \cite{moseley2016}).
The curves are closely mirrored by those of $r_{eff}$, where peaks also occur after approximately half the duration.
When rescaling both axes by the corresponding total track duration, all curves nearly collapse on one.
This indicates that both the peak track intensity as well as peak cell diameter are approximately proportional to track duration.

Also mergers show single-peaked life cycles (Fig.~\ref{fig:time_dependence_mergers}).
Precipitation intensity peaks are however generally higher and $r_{eff}$ already has appreciable values at the beginning of these tracks --- a feature that is not surprising, if it is considered that mergers are constituted by the concatenation of multiple solitary tracks at different, evolved, stages of their respective life cycles. As has already been discussed in \cite{moseley2016},  it is evident that both the initial precipitation intensity of mergers, as well as the initial value of $r_{eff}$ are elevated for the higher temperatures. 

\begin{figure}
    \centering
    \includegraphics[width=5.0cm,trim={0cm 0 0 0} ,clip]{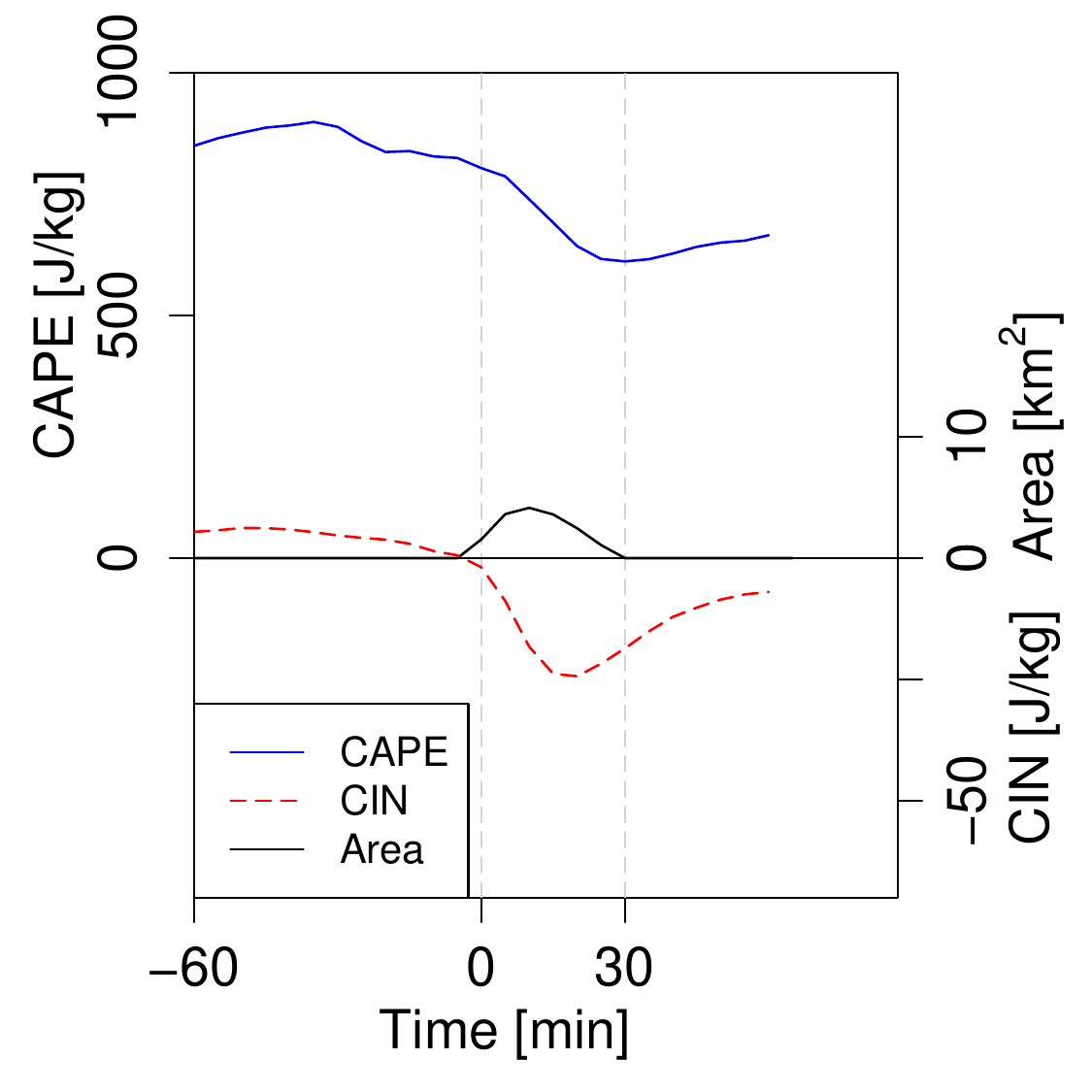}
    \includegraphics[width=5.0cm,trim={0 0 0 0} ,clip]{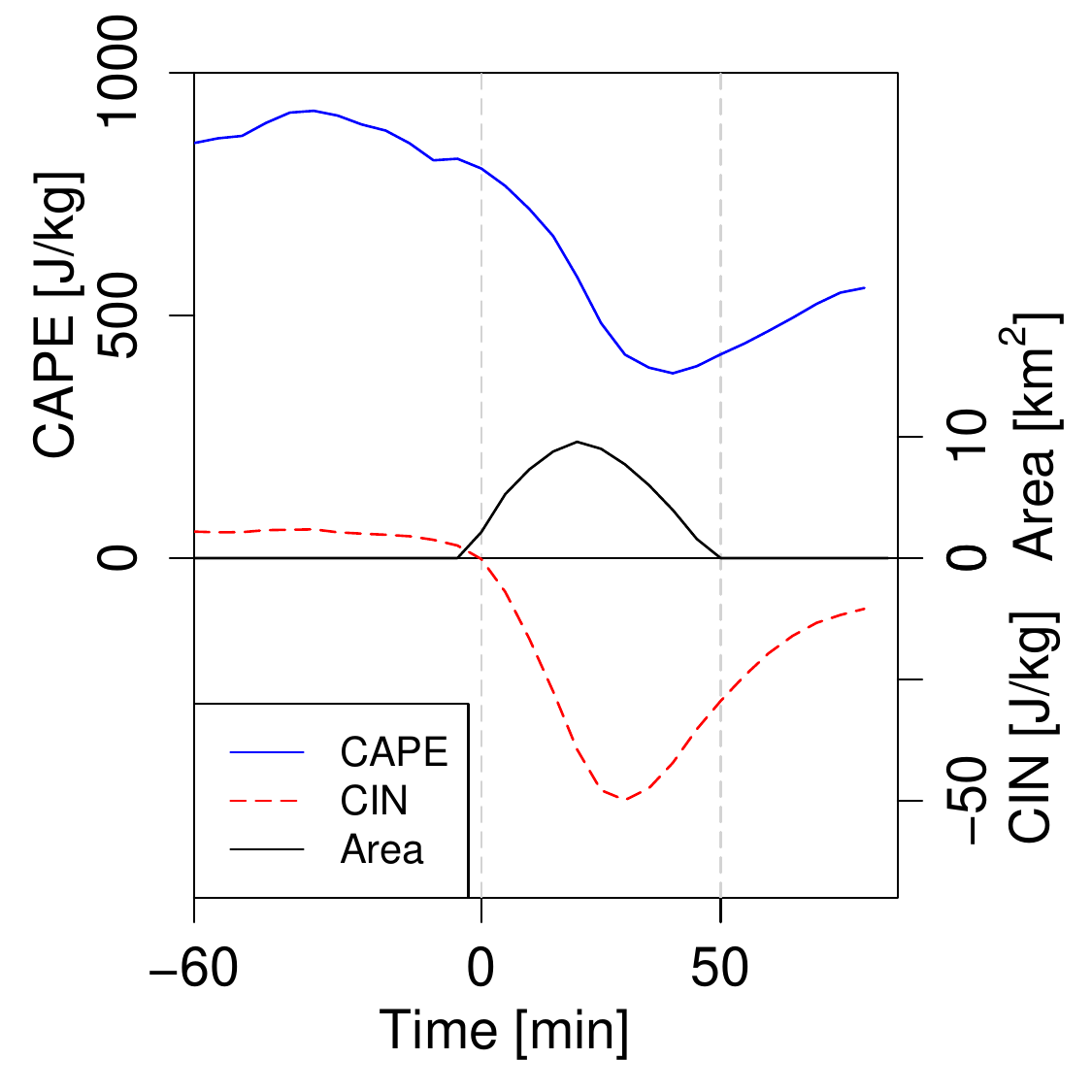}
    \caption{{\bf Mean CAPE life cycles.}
    Mean track area, CAPE, and CIN (defined as negative values), over all solitary tracks with 30 min (left) and 50 min (right) life time, for the P2K simulation and $\theta=1.0$. CAPE and CIN are given at the point of the track center of mass at the time of the maximum extent. Tracks begin at time $t=0$, negative values of $t$ show values before the onset of surface rain.
    }
    \label{fig:cape_timeseries}
\end{figure}

Also CAPE and CIN show very systematic behavior for purely solitary tracks (Fig.~\ref{fig:cape_timeseries}): Independent of the track lengths, CAPE usually originates at very similar values (near $800$ $J\,kg^{-1}$
, typical values in hurricanes and weak to moderate convection \citep{Jorgensen1989}) 
before precipitation onset and is only partially depleted within the track lifetime.
The reason for the fact that CAPE starts at similar values independent of the size that the tracks reach, is probably the similar atmospheric stratification for all events due to the absense of large scale forcing and advection.
Therefore, the view that is sometimes used in parametrizations of convection that CAPE is a measure of the total amount of precipitation produces, may be valid for averages over many single convective events, but not for each individual event: The total amount of CAPE before the onset of a single convective event cannot be used as a predictor for the strength, size, and duration that it will reach. 

In our results, CAPE rarely declines to zero after precipitation terminates (however, in some events this is indeed the case, see e.g. the example shown in Fig. \ref{fig:vpt_profiles}). 
The consumption of CAPE might be hampered by precipitation itself which builds up CIN by cooling the lower boundary layer and thus prevents air to rise further. 
This finding speaks to a partial relaxation, where some CAPE remains. Only after precipitation ceases, CAPE slowly begins to recover again. However, there is a clear relationship between the difference in CAPE at the beginning and the end of the tracks ($\Delta$CAPE) and track duration.

CIN, in turn, always originates near zero at precipitation onset and gradually builds up throughout the track duration (not shown) -- suggesting CIN$=0$ as a requirement for the onset of precipitation.
Shorter tracks -- those with smaller areas and intensities -- end up with lower CIN than the longer ones, possibly due to enhanced evaporative cooling within the boundary layer as a consequence of stronger rain evaporation.

Apart from the points mentioned above, we note that the definition of CAPE and CIN is somewhat ambiguous: 
an imaginary air parcel from within the boundary layer is lifted along a pseudoadiabat and CIN/CAPE are calculated in terms of the negative/positive buoyancy that this air parcel experiences in the surrounding atmosphere, according to the equations \ref{eq:cape} and \ref{eq:cin}. 
As a results, CAPE and CIN strongly depend on the level that the air parcel is lifted from. 
In our case, this is the lowest model level which is strongly cooled by the cold-pool that emerges as a result of evaporation rain. 
Therefore, a dominant reason for the rise in CIN and the depletion of CAPE immediately after the onset of surface rain is the marked drop in temperature in the lowest level caused by cold pools: 
a colder air parcel experiences more negative buoyancy (see the vertical profiles of virtual potential temperature $\theta_v$ for a selected solitary track in Fig. \ref{fig:vpt_profiles}). The other two effects that account for the reduction in CAPE during the track lifetime are the changes in the temperature profile inside the cloud due to latent heating, and due to mixing with the environment. 
These effects are however clearly small compared to the effect of boundary layer cooling. 
This stands in contrast to e.g. \cite{Zimmer2011} where it is assumed that tropospheric heating and boundary layer cooling and drying contribute approximately equally to the reduction of CAPE.

An overview of all four simulations CTR, P2K, P4K, and OMEGA is shown in Fig. S4, showing solitary track life cycles of intensity, temperature anomaly, and relative humidity. 
Again, the general picture is confirmed that for the three simulations CTR, P2K, P4K, i.e. in the absence of wind shear, solitary track life cycles are affected by the surface forcing at most weakly. Only the life cycles for OMEGA are clearly different, with weaker intensity peaks, and no significant anomalies in near-surface temperature and relative humidity, indicating that a CAPE/CIN analysis in this case is not possible any more.
However, in the absence of wind shear, the emergent cold pool is also clearly visible in the near surface temperature anomaly $\Delta T$, that is systematically depressed after the onset of rain events (Fig. \ref{fig:time_dependence_aux}). 
For longer-duration tracks, $\Delta T$ already recovers while precipitation is still ongoing --- an effect possible due to (dryer and therefore warmer) downdrafts caused by decaying convection.
We note that, at the end of the track, the temperature anomaly is nearly identical for all track durations, $\Delta T\approx -.4K$.
Relative humidity first increases (partially due to the decreases in temperature), but eventually becomes negative. 
The latter indicates a reduction also in specific humidity and could be explained by downdrafts, which bring relatively dry air down to the surface. 
This is partially attributable to the complication resulting from superposition of cells during the merging process.

\subsection{Linear dependencies}\label{sec:compressed_relations}

\begin{figure}
    \centering
    \includegraphics[height=4.5cm,trim={0 0cm 0 0},clip]{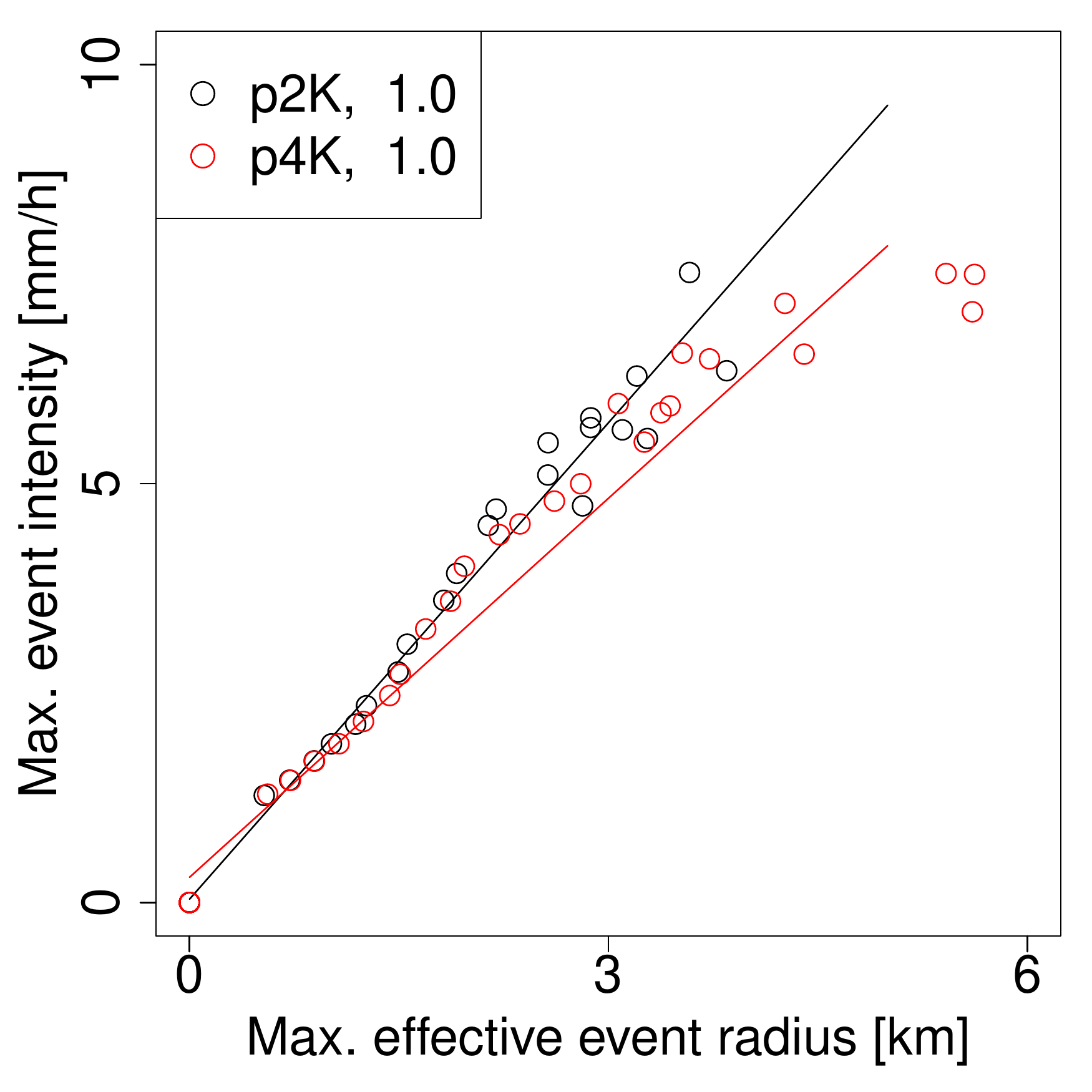}
    \includegraphics[height=4.5cm,trim={0 0cm 0 0},clip]{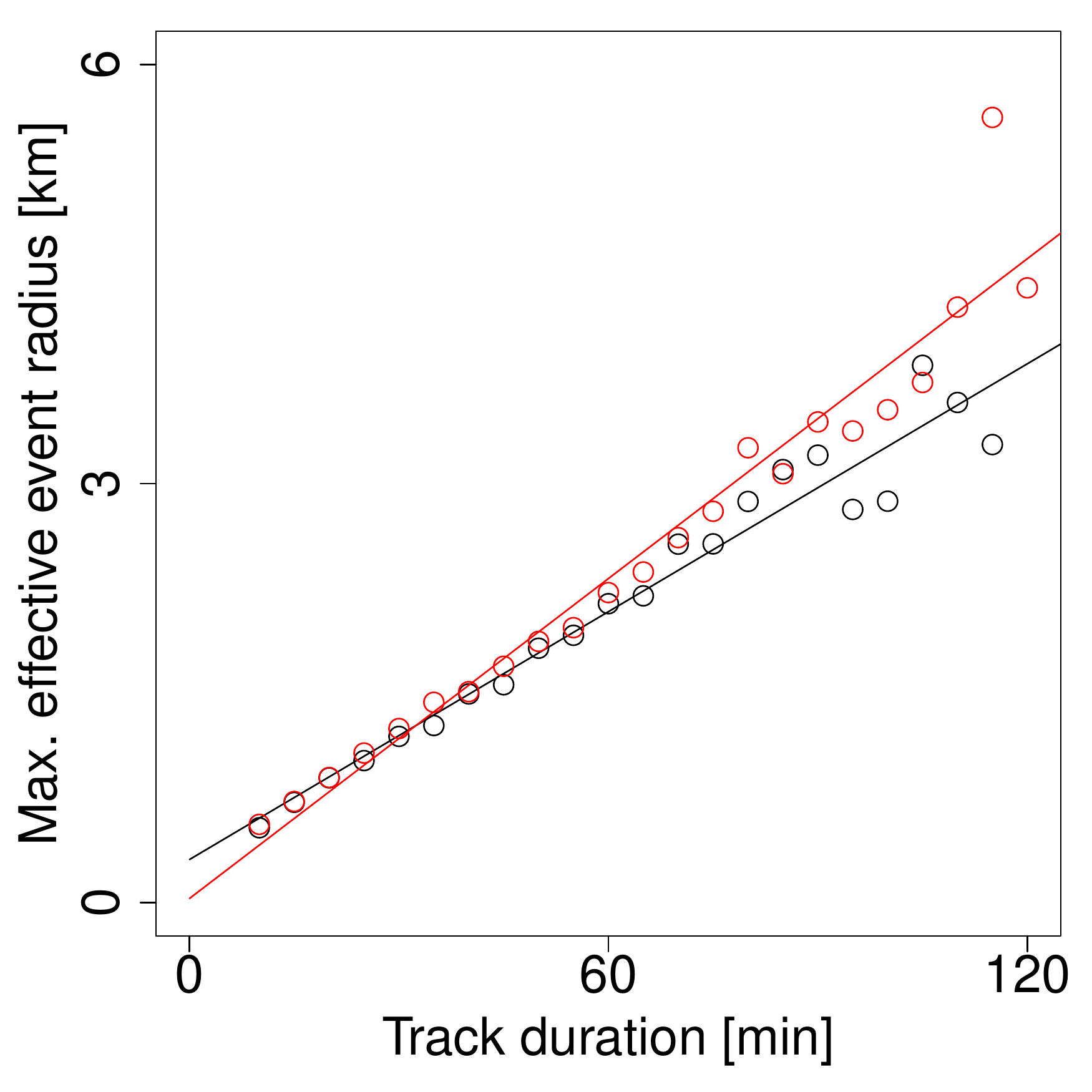}
    \includegraphics[height=4.5cm,trim={0 0cm 0 0},clip]{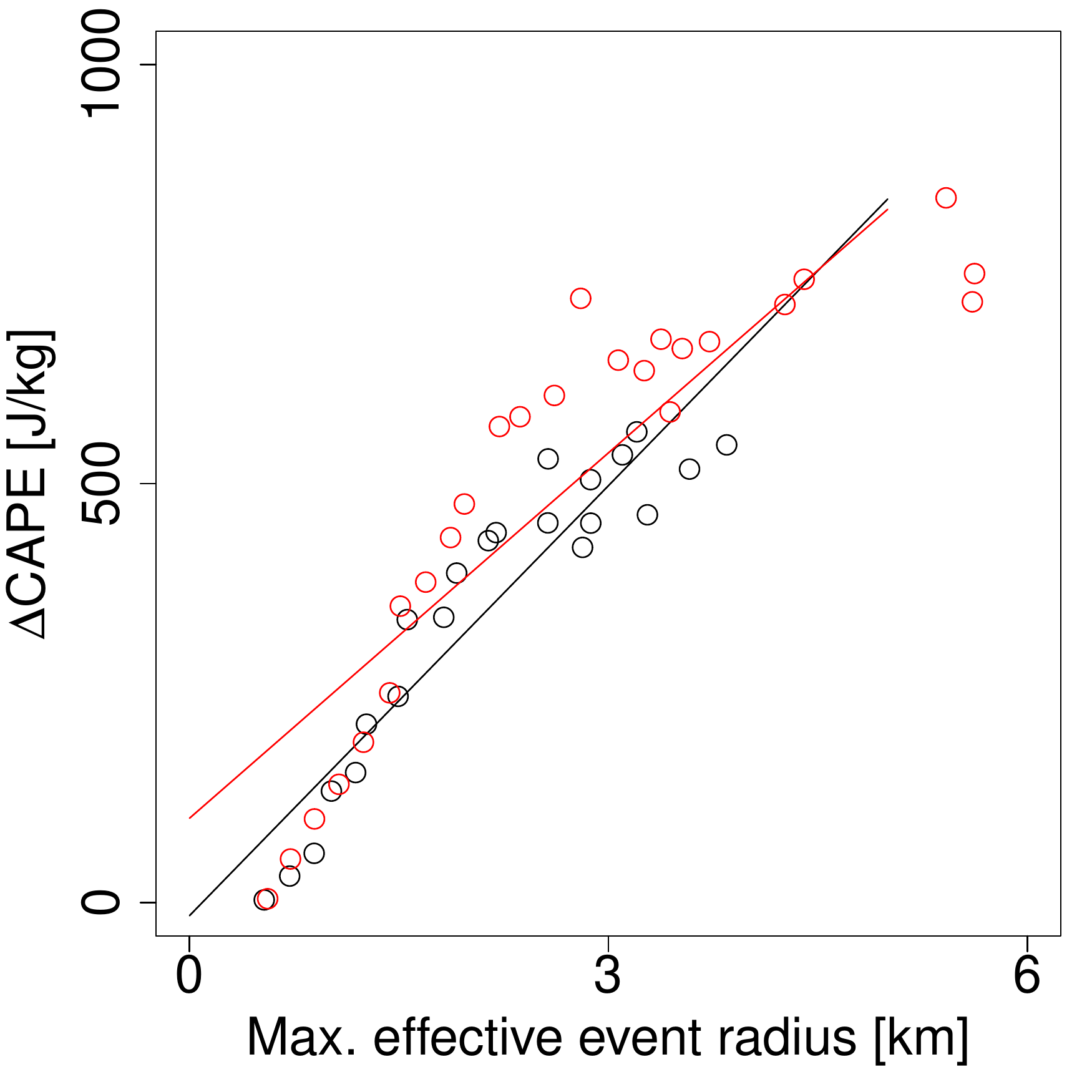}       
    \includegraphics[height=4.5cm,trim={0 0cm 0 0},clip]{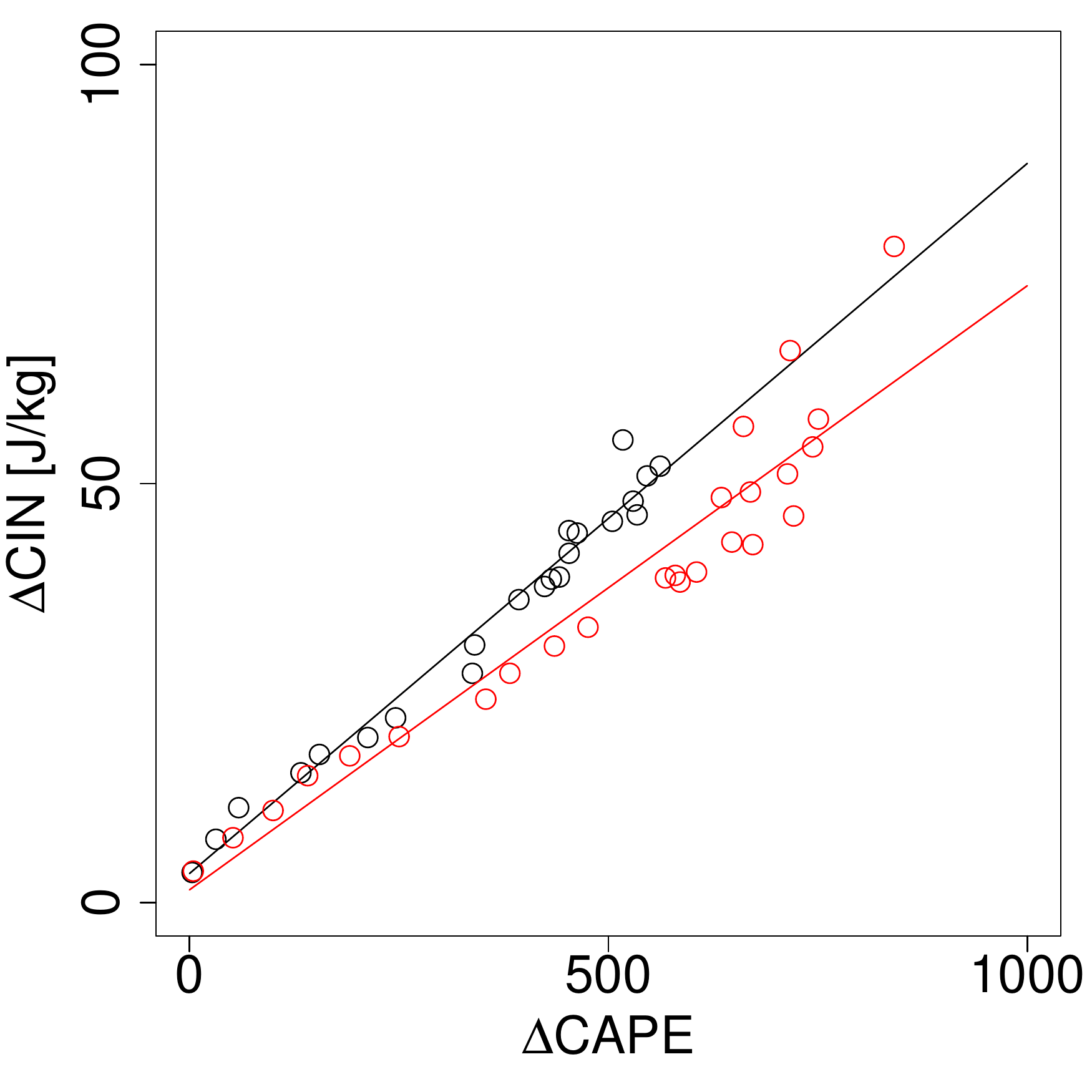}
    \includegraphics[height=4.5cm,trim={0 0cm 0 0},clip]{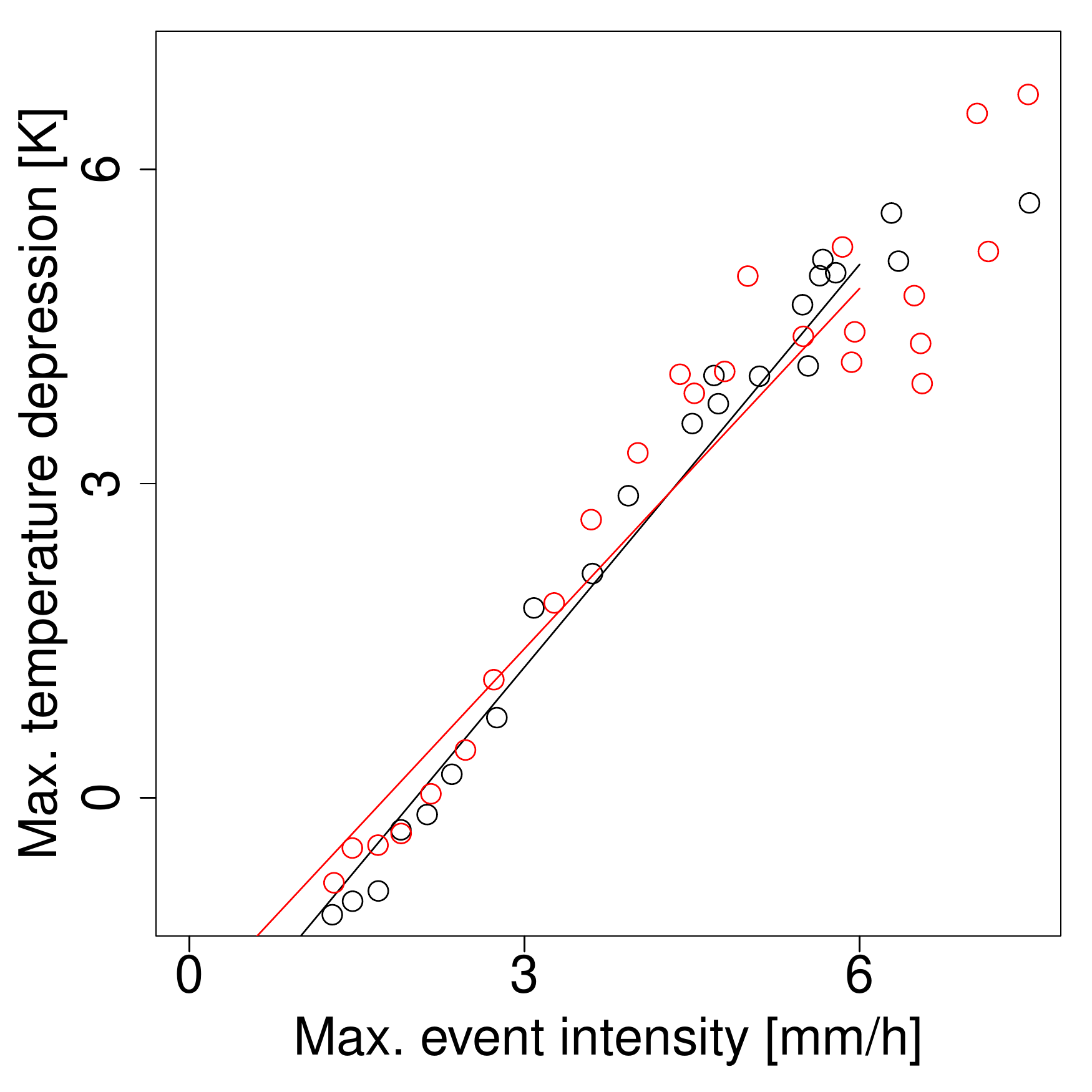}
    \caption{{\bf Various averaged relations for solitary tracks ($\theta=1$).} Solitary-solitary tracks for different parameters as shown in legend. Each circle shows a mean value averaged over all solitary tracks of a given duration. 
    Gray lines are linear fits to the data corresponding to the four parameter choices. 
    }
    \label{fig:various_relations}
\end{figure}

An interesting picture emerges when studying the relation between several quantities as they evolve throughout the tracks (Fig.~\ref{fig:various_relations}).
CAPE is transferred into lifting of air parcels as water vapor is condensed during ascent.
Considering that precipitation might indicate the reduction of CAPE due to latent heating,
we compare CAPE and maximum rain intensity, defined here as 
\begin{eqnarray}
I_{max}\equiv max(I(t), t_i\le t\le t_f)\;,
\end{eqnarray}
i.e. simply the maximum over the precipitation rate during the track lifetime.
Indeed, the relation between the two quantities is nearly linear, 
that is, the loss of CAPE roughly proportional to the accumulated precipitation amount. 
A similar linear relation holds for CIN. 
Hence, together, 
\begin{eqnarray}
\Delta CIN \sim \Delta CAPE \sim I_{max}\;,
\end{eqnarray}
As already discussed above, both CAPE and CIN react to rain evaporation, which effectively shifts boundary layer temperatures proportional to the mass of rain evaporated. 
Indeed, if the tropopause temperature were constant during the track lifetime, and a change in surface temperature were linearly relaxed all the way up to the tropopause, and further assuming a boundary layer height of $1\;km$ and tropopause height of $15\;km$, the change in CAPE would amount to 
\begin{eqnarray}
\Delta CAPE\approx g(h/2)\Delta T/T_{ref}\approx 200\;J/kg\;,
\end{eqnarray}
where $g=9.81ms^{-2}$ and $\Delta T\approx 1K$ is assumed.
However, longer tracks far exceed $\Delta CAPE\sim 200\;J/Kg$, which suggests that --- at least in those cases --- most of the change in potential energy is due to heating within the cloud level.

\begin{figure}
    \centering
    \includegraphics[height=4.5cm,trim={0 0 0 0},clip]{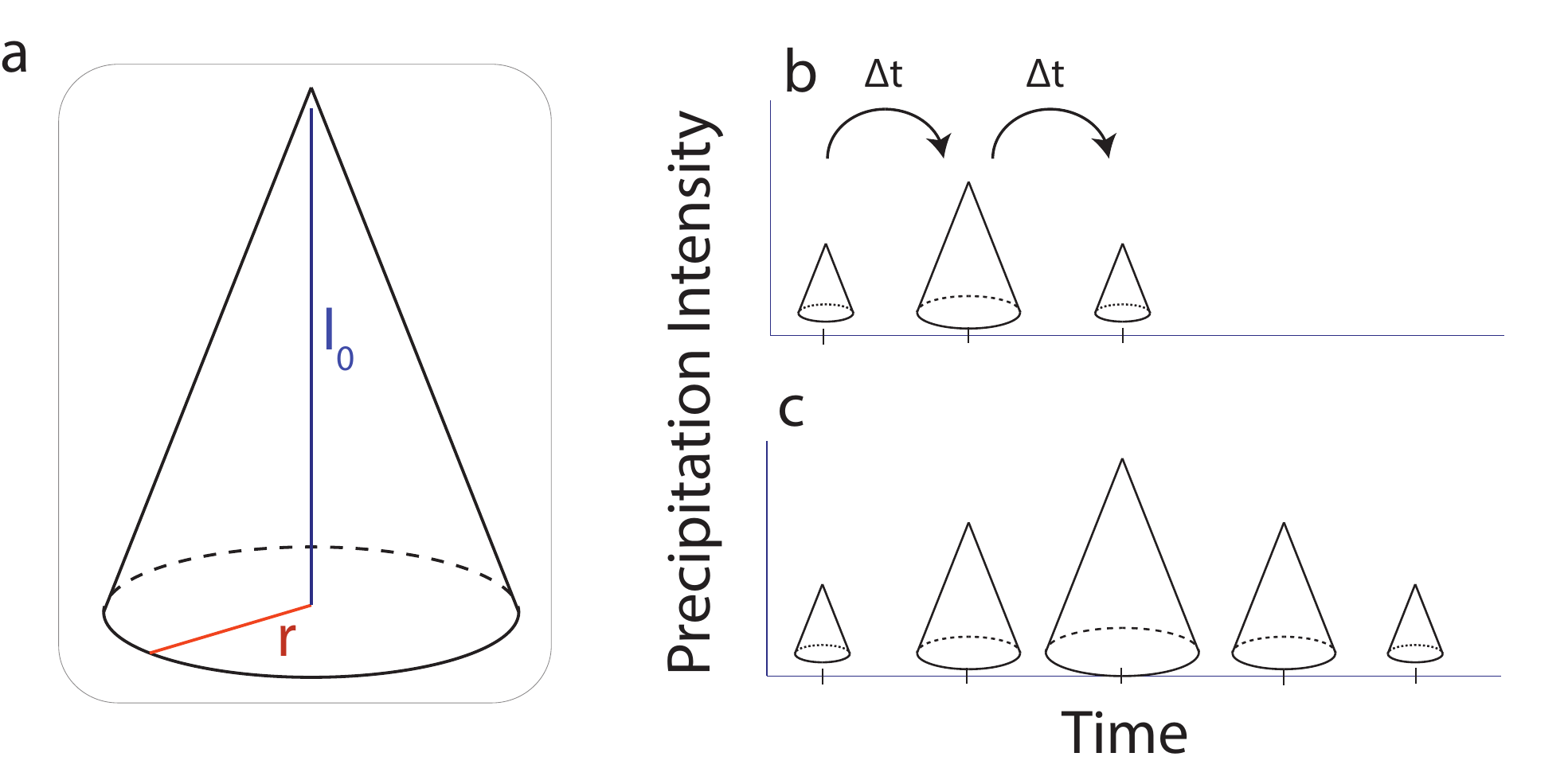}
    \caption{{\bf Cone model schematic.} 
    Cartoon image to the top right shows: (a) the simple model for track intensity vs. radius as a conal object in the space $(x,y,I)$; (b) an example of a short track; (c) a longer track. 
    }
    \label{fig:cone_schematic}
\end{figure}

When plotting peak intensity vs. track duration, we find that the peak scales roughly linearly with track duration: 
Short tracks of 10---20 minutes reach peak intensities of less than 2 $mm\,h^{-1}$ while tracks that last one hour reach peaks of more than 4 $mm\,h^{-1}$.
A geometric measure of patches is the maximum effective patch radius, defined as
\begin{equation}
    r_{max}^{eff}=\max_i{\{\sqrt{A_i/\pi}\}}\;,
\end{equation}
where $i$ enumerates all instances of the object within the track.
Our results show that maximum cell intensity again scales linearly with $r_{max}^{eff}$, implying that also cell radius and track duration are coupled linearly: longer duration tracks grow to proportionately larger radius, i.e. quadratically larger surface area.
Similar relations are consequently implied for the reduction in CAPE and increase in CIN, and for the depression of surface-near temperaure and relative humidity.

\noindent
\subsection{A statistical model for solitary tracks.} \label{sec:a_simple_model}

The previous discussion shows that solitary tracks have rather consistent properties, where the radius, intensity, and track lifetime are strongly linked.
Consider therefore a simple geometric model (Fig.~\ref{fig:cone_schematic}a---c) for a precipitation track, where cells have a circular cross section and the intensity peaks at the center of this circle (where its value is $I_0$) and decays linearly at a rate $\alpha$ that is similar for all cells, i.e.
\begin{eqnarray}
I(r)=I_0-\alpha r\; .
\label{eq:cone_model}
\end{eqnarray}
Further, cells are rotationally symmetric, i.e. there is no azimuthal dependence of $I$. 
The geometric structure formed by $(x,y,I)$ is hence a cone.
The maximum radius of a cell is then determined by $I(r)=0$, hence, $r_{max}=I_0/\alpha$.
Cell area $A(I_0)$, i.e. the cross section projected onto the surface, is 
\begin{eqnarray}\label{eq:cone_model_area}
A(I_0)=2\pi\,r_{max}^2=2\pi I_0^2/\alpha^2\;.
\end{eqnarray}
The spatially-averaged cell intensity becomes 
\begin{eqnarray}\label{eq:cone_intensity}
\overline{I}(I_0)&=&2\pi/A(I_0)\int_{r=0}^{r_{max}}dr\,rI(r)=I_0/6\,.
\end{eqnarray}
Hence, indeed, cells with larger cross section radius would produce proportionately larger average intensities, i.e. $\overline{I}\sim r_{max}$.
Further, knowing the average intensity, one also knows the maximum intensity -- and vice versa. 
This is in line with the findings in other studies \citep{Grabowski:2006,boing2012influence,schlemmer2014}.

\begin{figure}
    \centering
    \includegraphics[width=10cm,trim={0 0 0 0cm},clip]{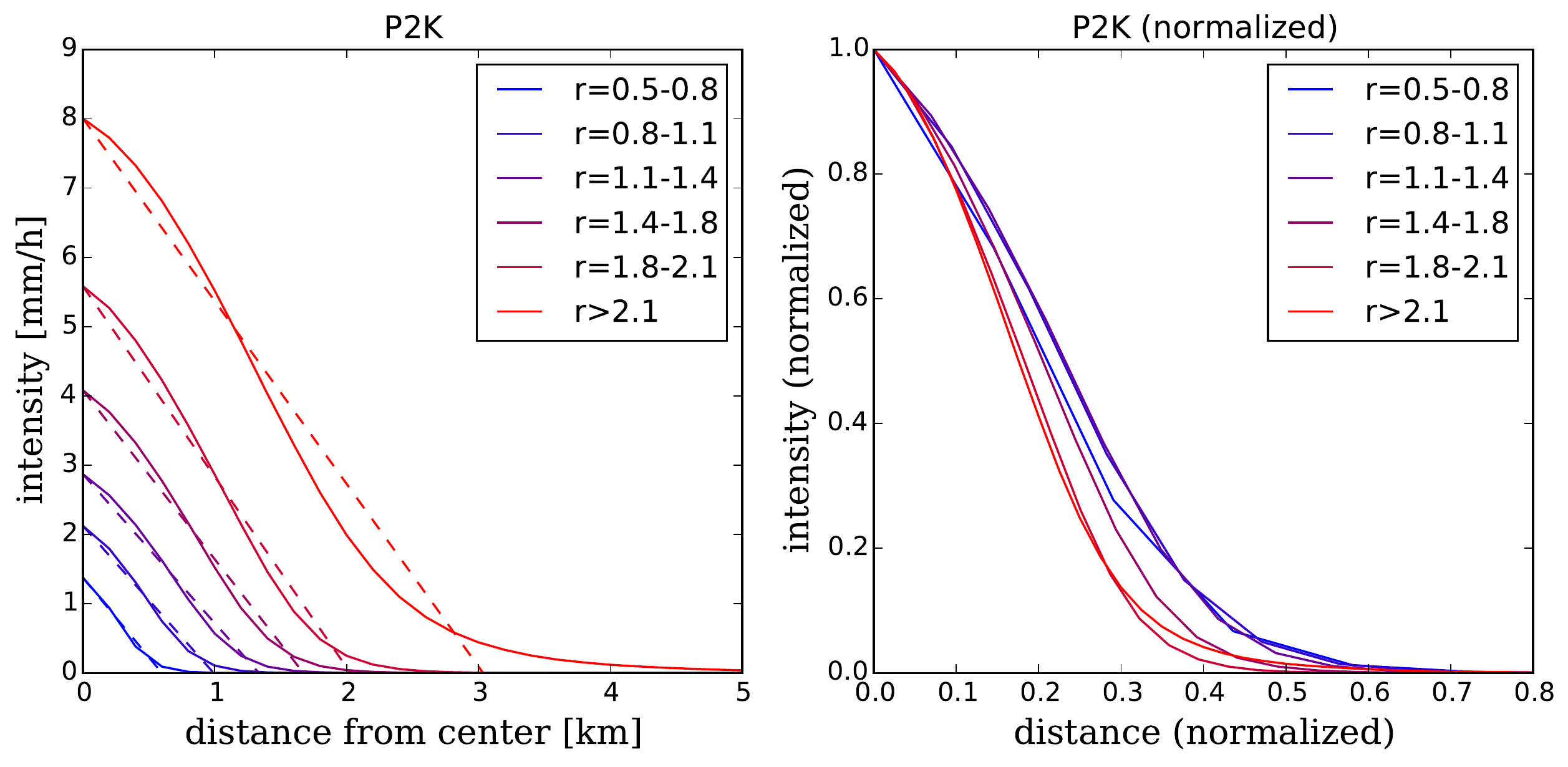}
    \caption{{\bf Radial precipitation intensity profiles.} Mean precipitation intensity as a function of distance from the object center of mass, averaged over composites of all objects with similar effective radius $r_{eff}$ (given in [km]), for the P2K simulation, and $\theta=1$. 
    {\it Left:} Absolute intensity versus distance from object mass center. Solid lines show averages over the object composites, dashed lines show fitted linear curves as given by Eq. \ref{eq:cone_model}. 
    {\it Right}: Analogous, but both axes scaled by the respective peak intensity, i.e. the intensity in the object center of mass.
    }
    \label{fig:radial_profiles}
\end{figure}

To check the validity of this simplified model, we build composites of objects of different areas. 
We average the intensity along circles around the center of mass of each object, while all grid boxes outside of each object's mask are assigned the value zero. 
Then, the resulting radial intensity profiles of all objects within the given area range are averaged. 
For solitary tracks, the resulting composite profiles are shown in Fig. \ref{fig:radial_profiles}. 
Even when the objects belonging to other track types are included, little difference in the results is seen (not shown). 
The explanation for the tails of the profiles of larger area objects is that a small number of objects have an elongated geometry, i.e. they strongly deviate from a circular shape and thus have non-zero intensity also at larger distances (see Fig.~\ref{fig:conemodel_sketch} for illustration).

The lines of the linear model as given by Eq.~\ref{eq:cone_model} shown in Fig.~\ref{fig:radial_profiles} are fitted such that the averaged cell intensity given by Eq.~\ref{eq:cone_intensity} is equal to the one calculated from the respective profiles, when $I_0$ is taken as the peak intensity at $r=0$. 
The fitted values for $\alpha$ are shown in Table \ref{table:alpha} and vary between $2.14$ and $2.75$ for P2K and $\theta=1$.
Even the differences between the composites of {\it CTR}, {\it P2K} and {\it P4K}, and between the choice of $\theta=0.5$, and $\theta=1$ are rather moderate. However, there is a tendency for smaller values of $\alpha$ for stronger surface forcing (i.e. for P4K, with respect to CTR and P2K), especially for larger cell sizes, which hints to more wide spread events in the case of higher forcing, probably due to the fact that more interaction happens since the distance between tracks is smaller. This asumption is supported by the fact that $\alpha$ is also smaller for $\theta=1$, compared to $\theta=0.5$, as in the former case the solitary tracks are more contaminated by merging of smaller tracks.

Thus, we conclude that our simple linear model proposed in this section works sufficiently well. Although an exponentially decaying profile was suggested from a radar analysis \citep{Hardenberg:2003}, the advantage of a linear profile is a clearly defined spatial extent of the rain cells.

\begin{figure}
    \centering
    \includegraphics[width=14cm,trim={0 0 0 0},clip]{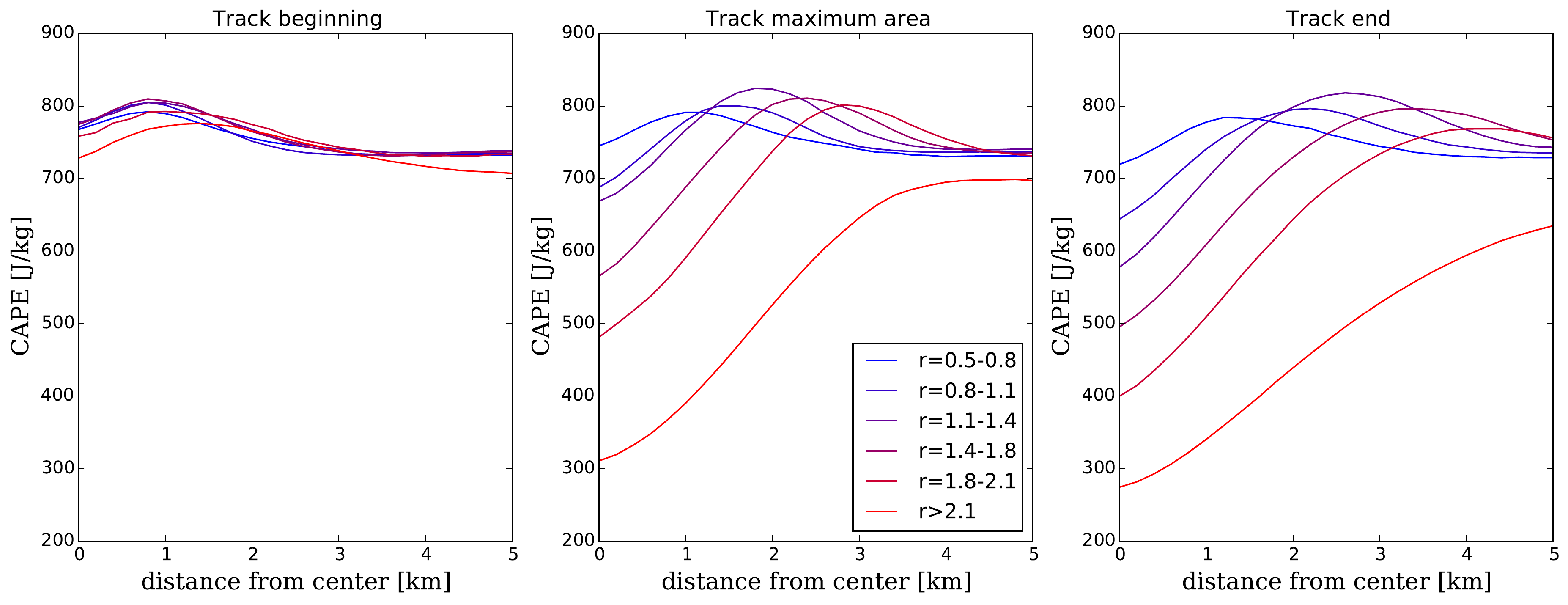}
    \caption{{\bf Radial profiles of CAPE.} Mean CAPE as a function of distance from the object center of mass, averaged over composites of all tracks with similar maximum effective radius $r_{eff,max}$, for the P2K simulation, and $\theta=1$. 
    {\it From left to right:} Mean profile at the initial time step of each track, at the time when each track reached its maximum extent, and at the last time step of each track.}
    \label{fig:radial_profiles_cape}
\end{figure}

To shed more light on the spatial distribution of CAPE during the course of the tracks, in a similar way we show composits of radial CAPE profiles, averaged over the beginning, the time of maximum extent, and the end of each solitary track (Fig. \ref{fig:radial_profiles_cape}), again conditioned on tracks of similar maximum effective radius $r_{eff}$. As a supplement to the time series shown in Fig. \ref{fig:cape_timeseries}, these radial profiles confirm that the original CAPE is only weakly disturbed at the onset of rain events, but then reaches a strong depletion at the time of maximum track extent, especially toward the track center. When rain ceases, this strong depletion still remains and grows in extent.

\begin{table}
 \caption{Fitted values for $\alpha$ from the curves shown in Fig. \ref{fig:radial_profiles} in units of $mm\;h^{-1}km^{-1}$}\label{table:alpha}
\begin{center}
\begin{tabular}{c||c|c|c|c|c|c} 
\hline
 $r_{max}$ in $[km]$  & 0.5--0.8 & 0.8--1.1  & 1.1--1.4 & 1.4--1.8 & 1.8--2.1 & $>$2.1     \\ \hline\hline
         CTR ($\theta=0.5$)  & 2.45      & 2.40        & 2.58       & 2.98       & 3.22       & 3.48        \\ \hline
         CTR ($\theta=1.0$)  & 2.46      & 2.37        & 2.53       & 2.93       & 3.15       & 3.38        \\ \hline
         p2K ($\theta=0.5$)  & 2.31      & 2.17        & 2.20       & 2.50       & 2.81       & 2.87        \\ \hline
         p2K ($\theta=1.0$)  & 2.29      & 2.14        & 2.15       & 2.44       & 2.75       & 2.64        \\ \hline
         p4K ($\theta=0.5$)  & 2.81      & 2.05        & 2.22       & 2.42       & 2.58       & 2.62        \\ \hline
         p4K ($\theta=1.0$)  & 2.30      & 2.02        & 2.15       & 2.34       & 2.54       & 2.21        \\ \hline
\end{tabular} 
\end{center}
\end{table}




\section{Conclusion and Outlook}\label{sec:discussion}

We have shown that a simple rain cell tracking algorithm can elucidate a number of track properties and their relations. 
The tracking allows us to isolate the dynamics of tracks that begin and end autonomously, that is, without interference such as merging or fragmentation.
To avoid the termination of larger tracks that interact with smaller ones, we introduced a parameter $\theta$ that controls merging and splitting, and show that the statistics of such solitary tracks relatively robust with respect to $\theta$.
For the solitary tracks, we summarize our findings in a simple statistical model with spherical symmetric cells where the intensity falls of linearly with distance from the center (Eq.~\ref{eq:cone_model}). 
This model relates precipitation area, track lifetime and precipitation intensity and gives a reasonable fit for solitary events of all sizes. 
Our findings imply that much of the initiation and termination of these isolated convective cells may be driven by processes occurring within the boundary layer --- in particular, termination of convective updrafts may be mainly caused by cooling through rain re-evaporation, rather than reductions in cloud level buoyancy.

It is remarkable, that, at least for the simple LES modeling set-up without wind shear and homogeneous surface characteristics, solitary tracks follow systematic and relatively simple relations between the different quantities.
This finding is encouraging, since, in this simple setup, there is hope for compressing the information on the cloud cell population into only few parameters. In particular, we have discussed the relation between the drop in CAPE, the drop in surface-near temperature and humidity due to rain evaporation, and the intensity that a track reaches. This indicates that, at least for a mean description of convective lifecycles, one of these parameters is sufficient for our simple statistical model, as they are all linearly interrelated. However, for an explanation of higher moments of the distribution of solitary track durations, more parameters might be requires, e.g. the total water content that has been identified as a crucial variable affecting moist convection by \cite{derbyshire2004}. 

Future extensions of the current study should address the distribution of maximum intensities, or perhaps equivalently, the life times of tracks. 
Further, we leave the question open if this picture could partially be carried over to merged tracks: 
When cells merge, they grow larger and produce stronger precipitation intensities.
Extreme convective precipitation may be a result of multiple, sequential merging incidents.
These tracks ({\it m-m} type) are statistically rare, but form the most intense precipitation, stemming from the largest cells with longest lifetimes.
Incidentally, these {\it m-m} tracks intensify and grow further, as surface forcing is increased.
It has been suggested that merging can invigorate precipitation intensity in tropical convective cells \citep{glenn2017connections}, as the interior of a merging object may be shielded against entrainment. 
Clouds can then become deeper due to the reduced entrainment and develop larger updraft velocities.
Our results show that tracks resulting from merging indeed lead to larger and more intense rainfall events with extended lifetimes. However, out study leaves the question open if the intensification is indeed {\it caused} by the merging process that can be explained, e.g. by a "screening" effect where the interior of a merged cell suffers less from mixing processes, or if it is merely a statistical effect resulting from the fact that larger cells have a higher probability to merge into others. In how far merging can be seen as a simple superposition, has to be clarified in a future study.

Another crucial process in convective organization, not addressed here, is explicit triggering by the cold pool dynamics, a topic currently heavily discussed in the community.
Cold pool interactions, in distinction to merging processes, likely occur at larger spatial separation of the original precipitation cells that cause the cold pools, and it has been suggested that the scales evolving throughout the convective diurnal cycle may increase systematically -- a possible consequence of the interplay of cold pool interaction and increasing convective inhibition (CIN).
Revisiting Fig.~\ref{fig:trackmask}, cold pools are responsible for the clearing of several subareas of the model domain, most visible at $t-t_0=5h$ for $T_0=25^{\circ} C$.
Such cold pool dynamics additionally increases the complexity of interactions and depletes the space available for precipitation cells to grow.

To give an outlook, future research in this area might benefit from a focus on processes facilitating repeated merging.
Fortunately, such processes, which often involve larger (6---8 $km$ diameter) and longer lasting (30---40 $min$) tracks, may be measurable over vast regions of the globe through satellite data, which often can reach resolutions as high as $1.5$---$3$ $km$ spatially, and $\sim 15$ $min$ temporally. 

\acknowledgments
C.M. acknowledges funding through the German Federal Ministry of Education and Research for the project "HD(CP)$^2$ - High definition clouds and precipitation for advancing climate prediction" within the framework programme "Research for Sustainable Development (FONA)", under the number 01LK1506F.
J.O.H. and O.H. acknowledge funding through a Grant (13168) by the Villum Foundation. 


 \newcommand{\noop}[1]{}

\clearpage

\renewcommand{\thesection}{S\arabic{section}}
\setcounter{section}{0}
\renewcommand{\thefigure}{S\arabic{figure}}
\setcounter{figure}{0}
\renewcommand{\theequation}{S\arabic{equation}}
\setcounter{equation}{0}

\title{Extreme convective precipitation as the result of a multi-merge selection process\\--- Supplementary Information  ---}\label{sec:supp}
\noindent
This supplementary information contains further text and figures that are not crucial for the understanding of the main text. However, this material may be useful for those readers who are interested in the details, which were mainly summarized in the text in the main manuscript.

\subsection{The effect of $\theta$}\label{sec:theta}

In Fig. \ref{fig:threshold_ratio} we give a general overview of the different track types and how their weight is affected by the parameter $\theta$.
As might be expected, the total precipitation contribution of solitary tracks monotonically increases for increasing $\theta$ (Fig. \ref{fig:threshold_ratio}, from ca. 25\% to ca. 75\% for {\it P2K}, and from ca. 10\% to ca. 75\% for {\it P4K}), while the contribution of mergers monotonically decreases. 
At $\theta=0$ the contribution of mergers is 32\% for {\it P2K} and somewhat larger (42\%) for {\it P4K}. 
At $\theta=1$, the contribution of the mergers is essentially zero, as identical areas of the merging or fragmented cells would be required. 
The track that is treated as continued is always the merging or fragmentation result, respectively, and therefore no new track initialization happens. 
The contribution of fragmentation also decreases with $\theta$, but is generally small. 
Also the contribution of other track types monotonically decreases with $\theta$, but remains finite even at $\theta=1$.
For values of $\theta$ between 0 and $0.5$ the main contribution is clearly from solitary tracks and mergers, therefore we will, in the following, often focus on those two track types.

We mention a few minor technicalities: 
Tracks that are only one time step long are neglected, however, in Fig. \ref{fig:threshold_ratio} their contribution is included into the track type {\it other}. 
Their contribution decreases somewhat when varying $\theta$ from 2.3\% to 0.3\% in {\it P2K} and from 5.8\% to 0.4\% for {\it P4K}. 
However, even at $\theta=1$ their contribution is still non-zero. 
For instance, it may occasionally be the case that an object splits off from one track and merges into another, larger track at the next time step. 
Such an object would initiate a track that would be terminated immediately and therefore would be only one time step long, track at $\theta=1$.


\begin{figure}
    \centering
    \includegraphics[width=14cm,trim={0 2.0cm 0 0},clip]{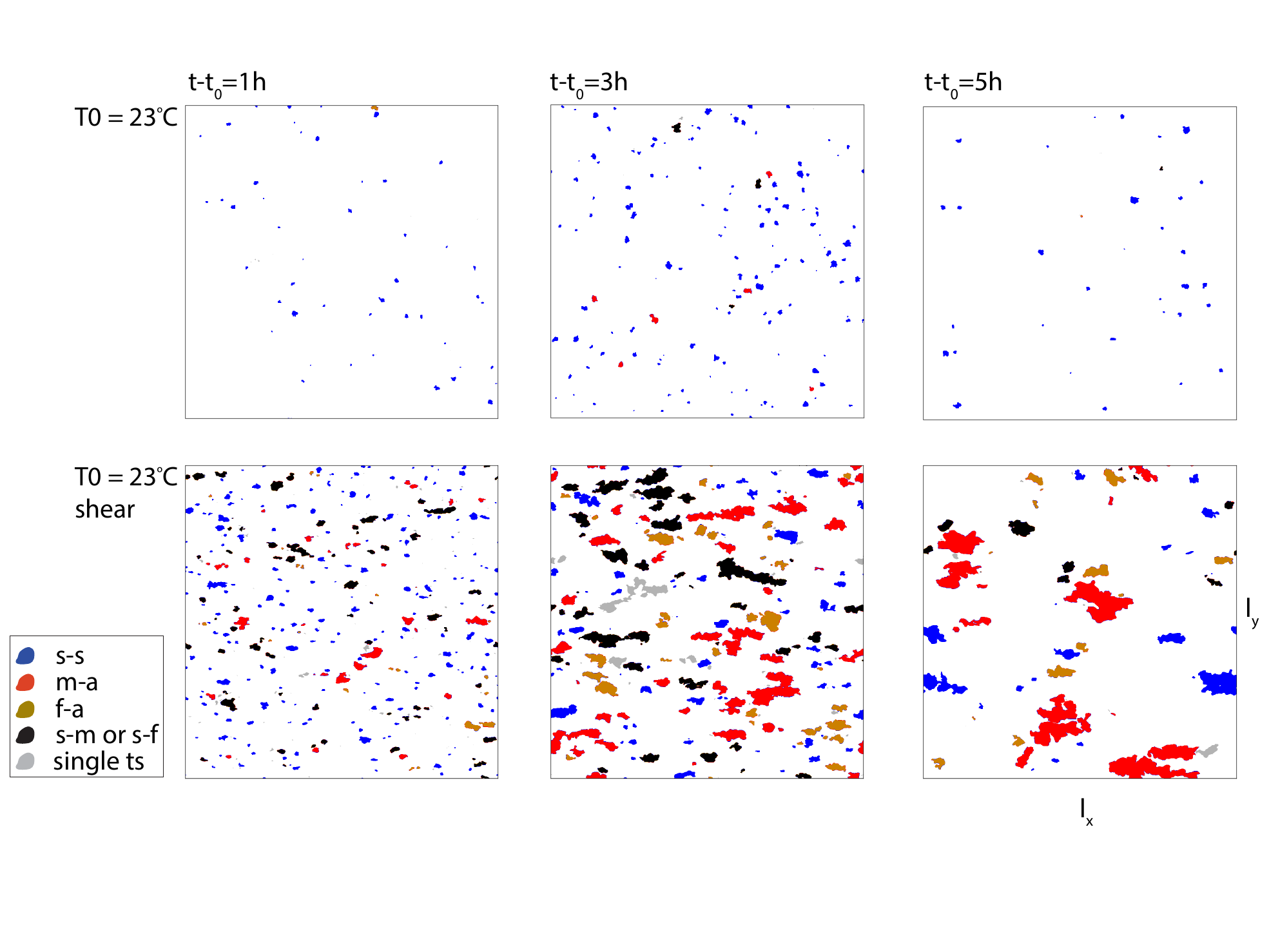}
    \caption{{\bf Time sequence of tracked objects.} Precipitation objects one hour, three hours and five hours after the onset of precipitation for $T_0=23^{\circ}C$ without wind shear ({\it CTR} simulation, upper row) and $T_0=23^{\circ}C$ with large-scale wind shear ({\it OMEGA} simulation, lower row), as labeled, for $\theta=0.5$. 
    Legend, coloring and remaining model parameters as in Fig.~\ref{fig:trackmask}.
    Objects are colored by their track type as indicated in the legend. 
    Note the pronounced merging effect caused by large-scale advection.
    }
    \label{fig:trackmask_supp}
\end{figure}

\begin{figure}
    \centering
    \includegraphics[width=3.6cm,trim={0cm 2.35cm 0 0} ,clip]{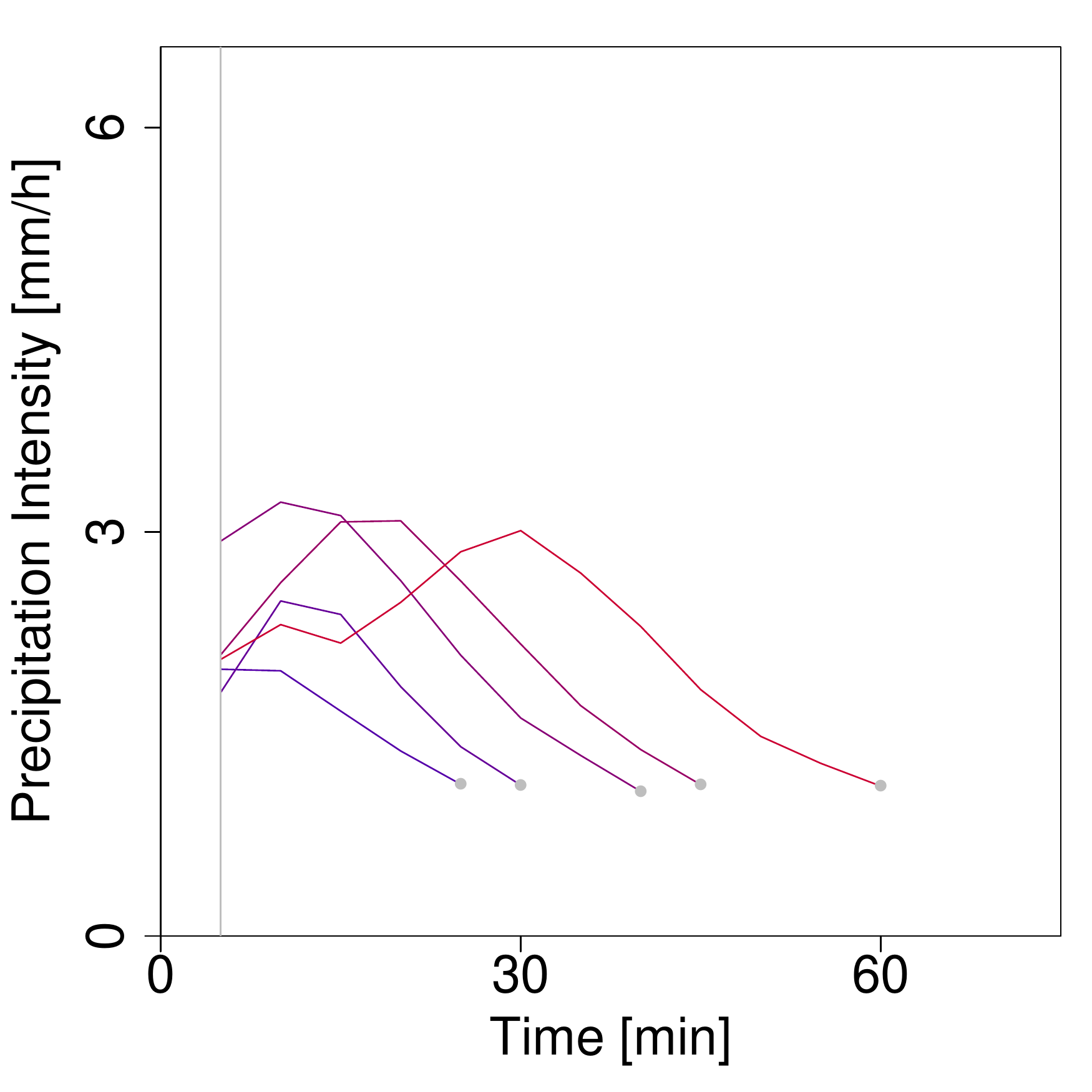}
    \includegraphics[width=3.2cm,trim={2.1cm 2.35cm 0 0} ,clip]{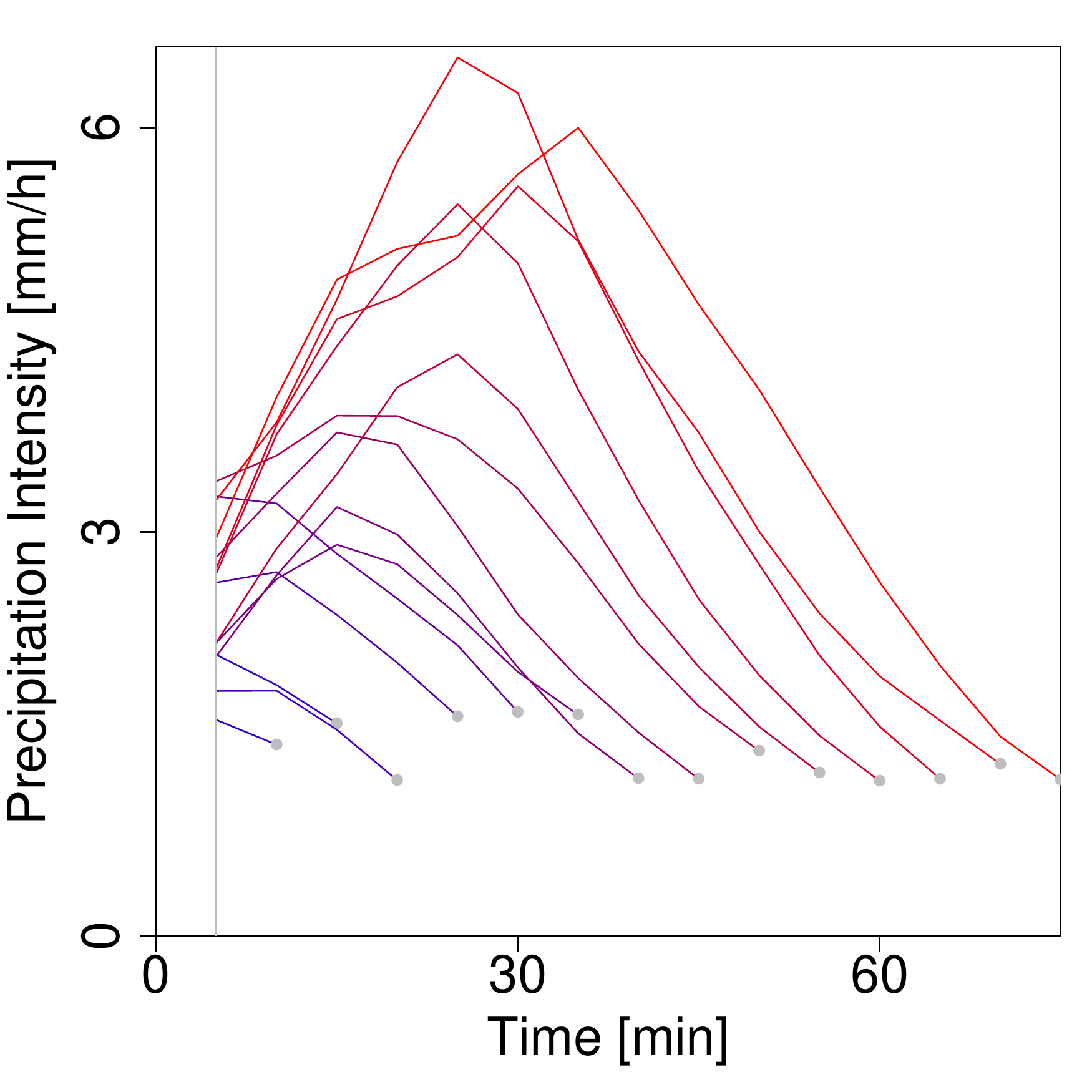}
    \includegraphics[width=3.2cm,trim={2.1cm 2.35cm 0 0} ,clip]{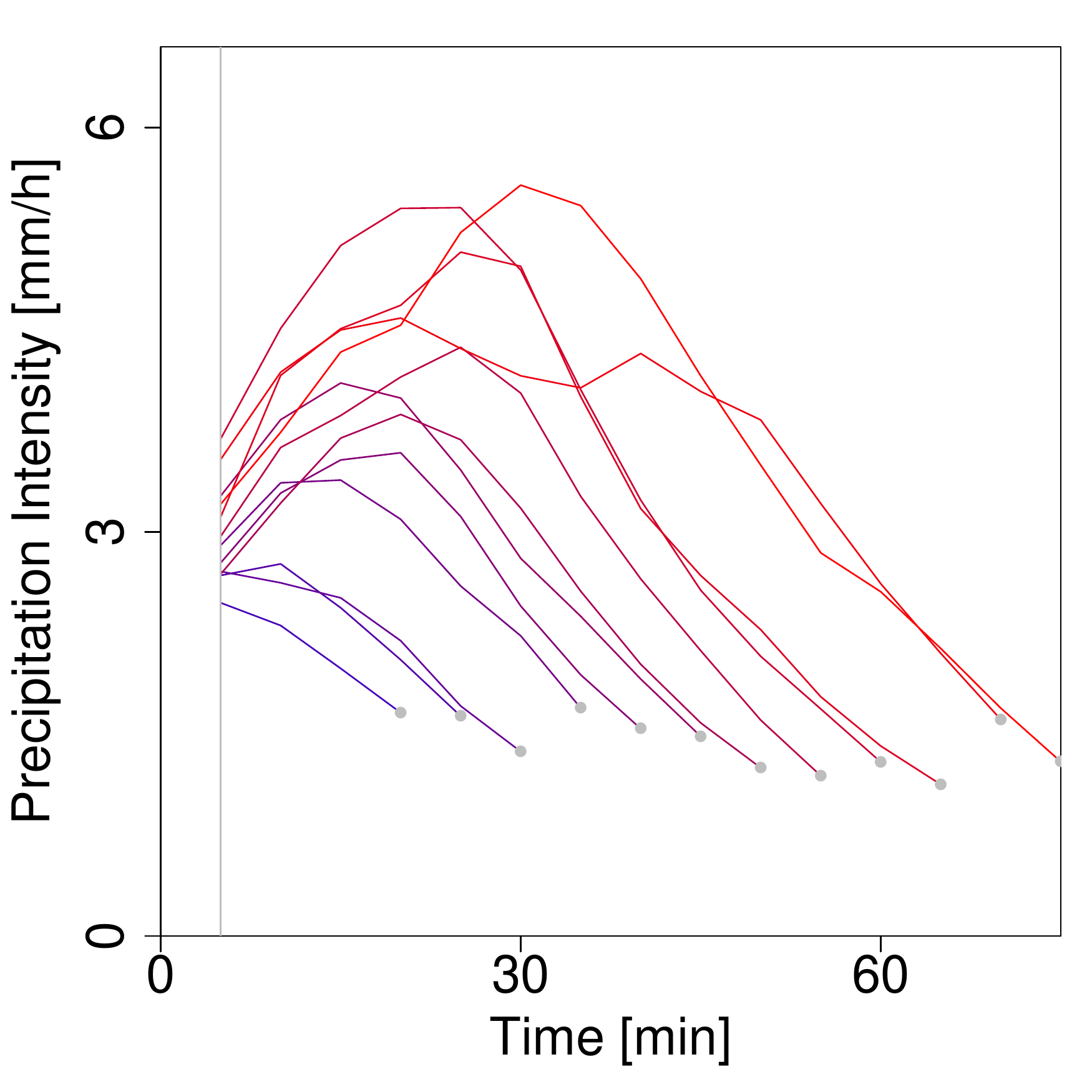}
    \includegraphics[width=3.2cm,trim={2.1cm 2.35cm 0 0} ,clip]{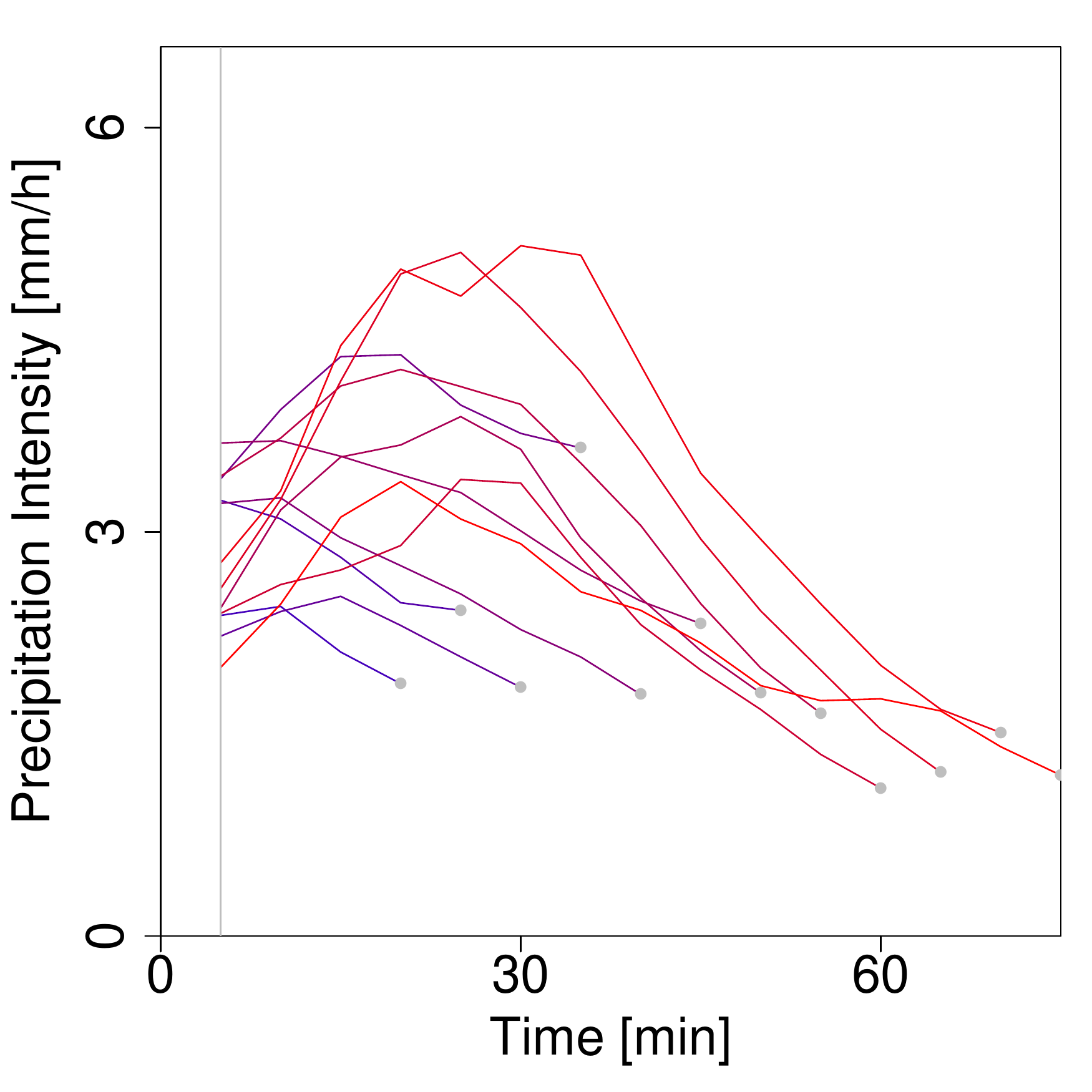}\\
    \includegraphics[width=3.6cm,trim={0cm 0 0 0},clip]{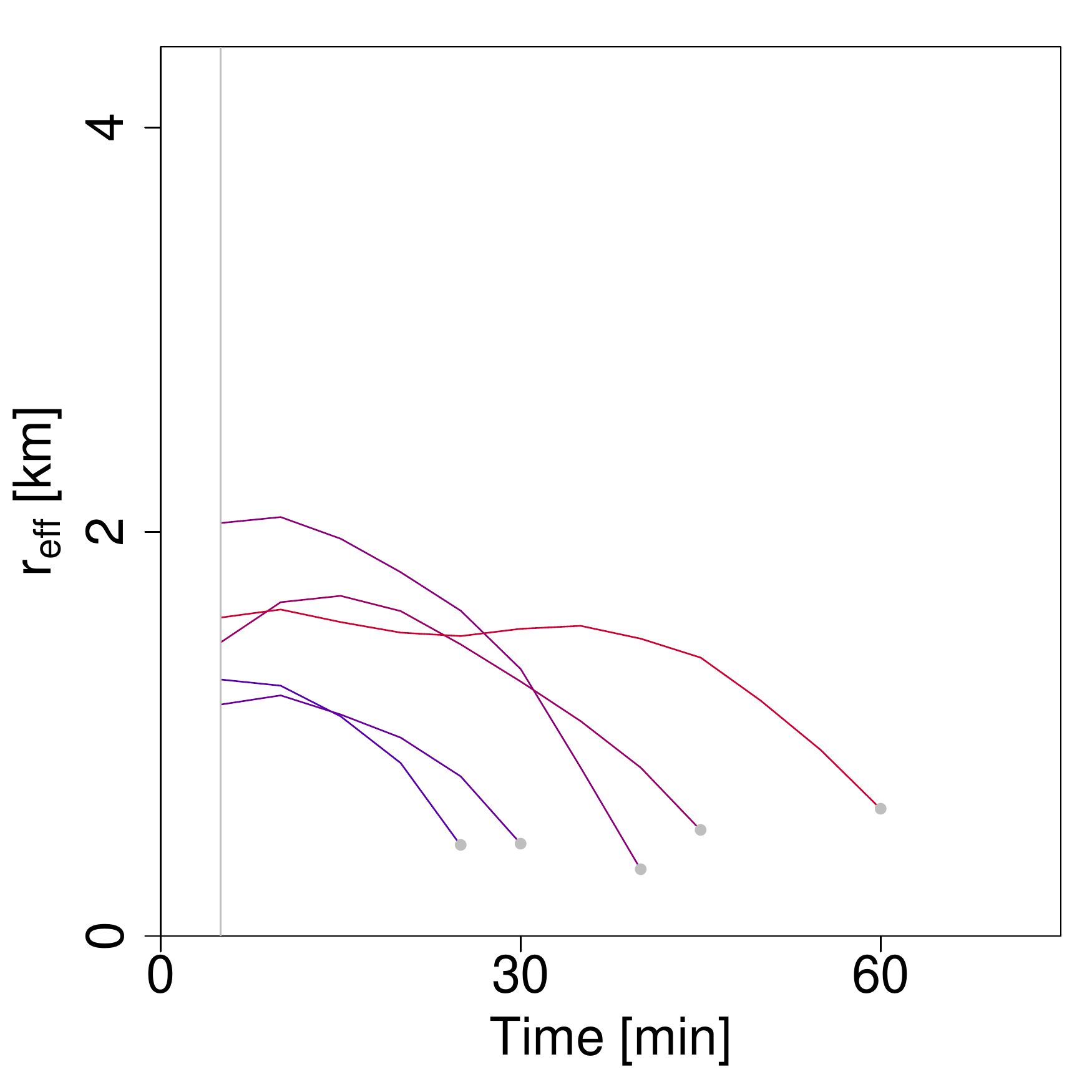}
    \includegraphics[width=3.2cm,trim={2.1cm 0 0 0},clip]{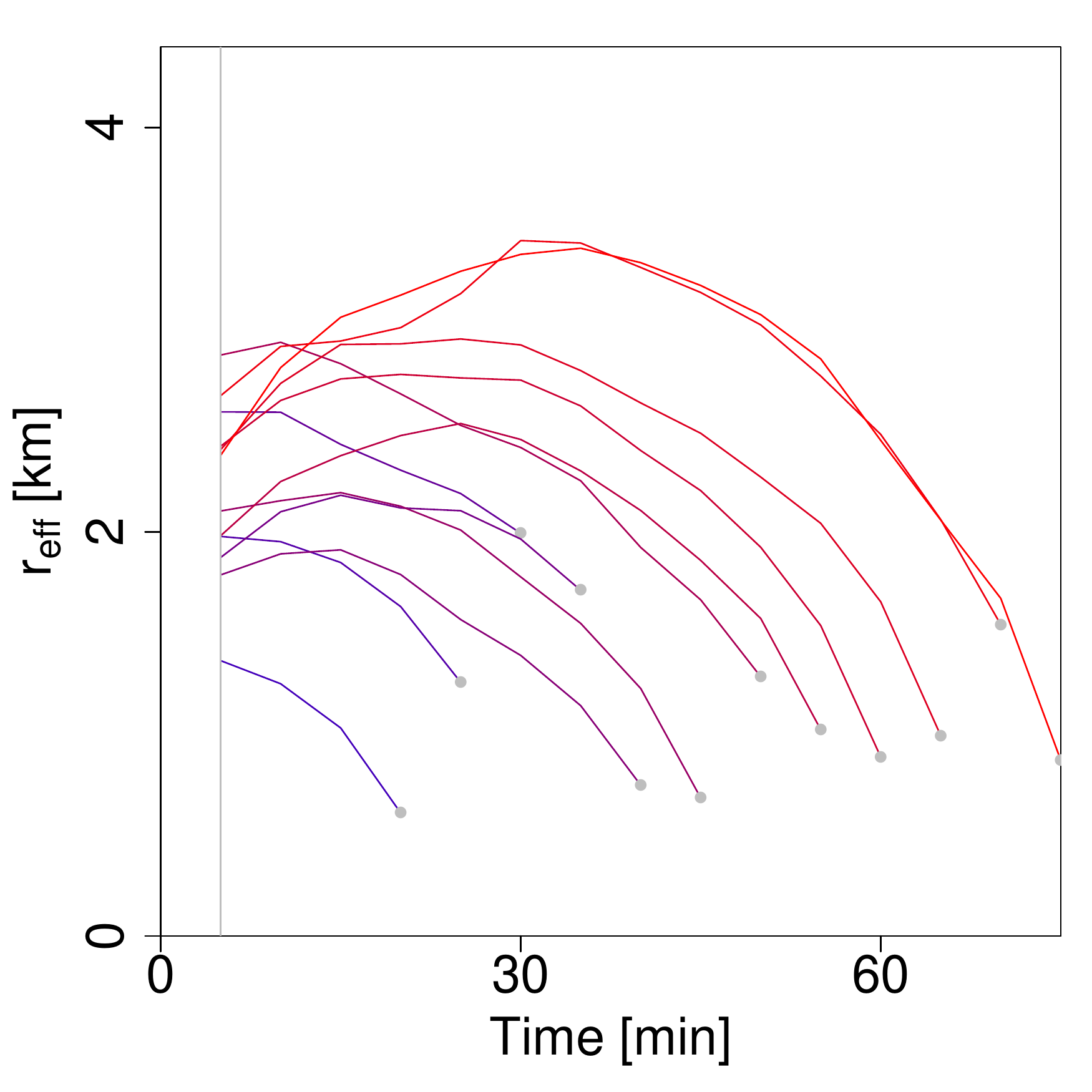}
    \includegraphics[width=3.2cm,trim={2.1cm 0cm 0 0},clip]{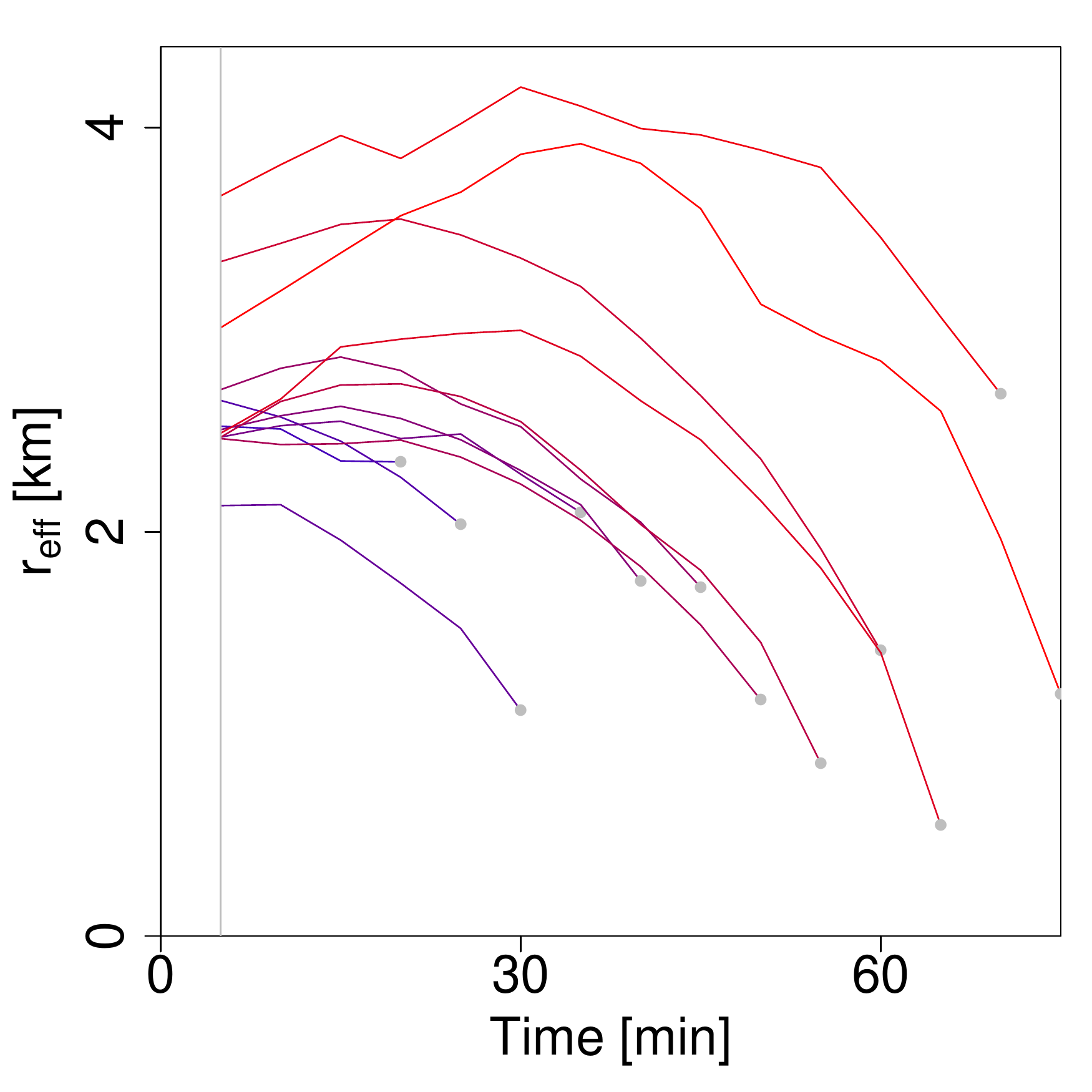}
    \includegraphics[width=3.2cm,trim={2.1cm 0cm 0 0},clip]{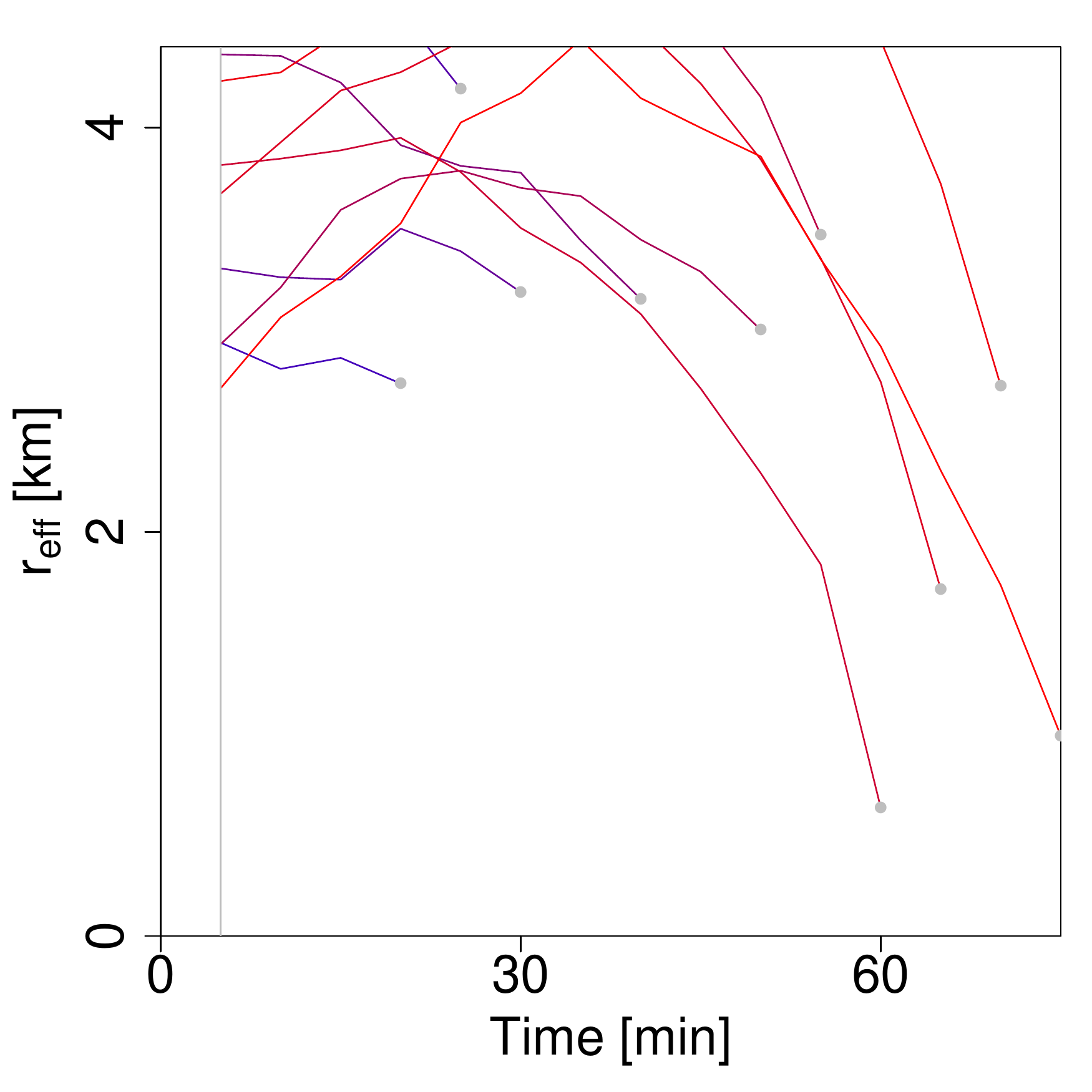}\\
    \caption{{\bf Time dependence across track duration (mergers).}
    Similar to Fig.~\ref{fig:time_dependence} but for tracks initiated as mergers (track types {\it m-a}): Intensity (top row), and effective radius (bottom row). Columns from left to right: CTR, P2K, P4K, and OMEGA simulation, each with $\theta=0.5$.
    }
    \label{fig:time_dependence_mergers}
\end{figure}

\begin{figure}
    \centering
    \includegraphics[width=6.0cm,trim={0 0 0 0} ,clip]{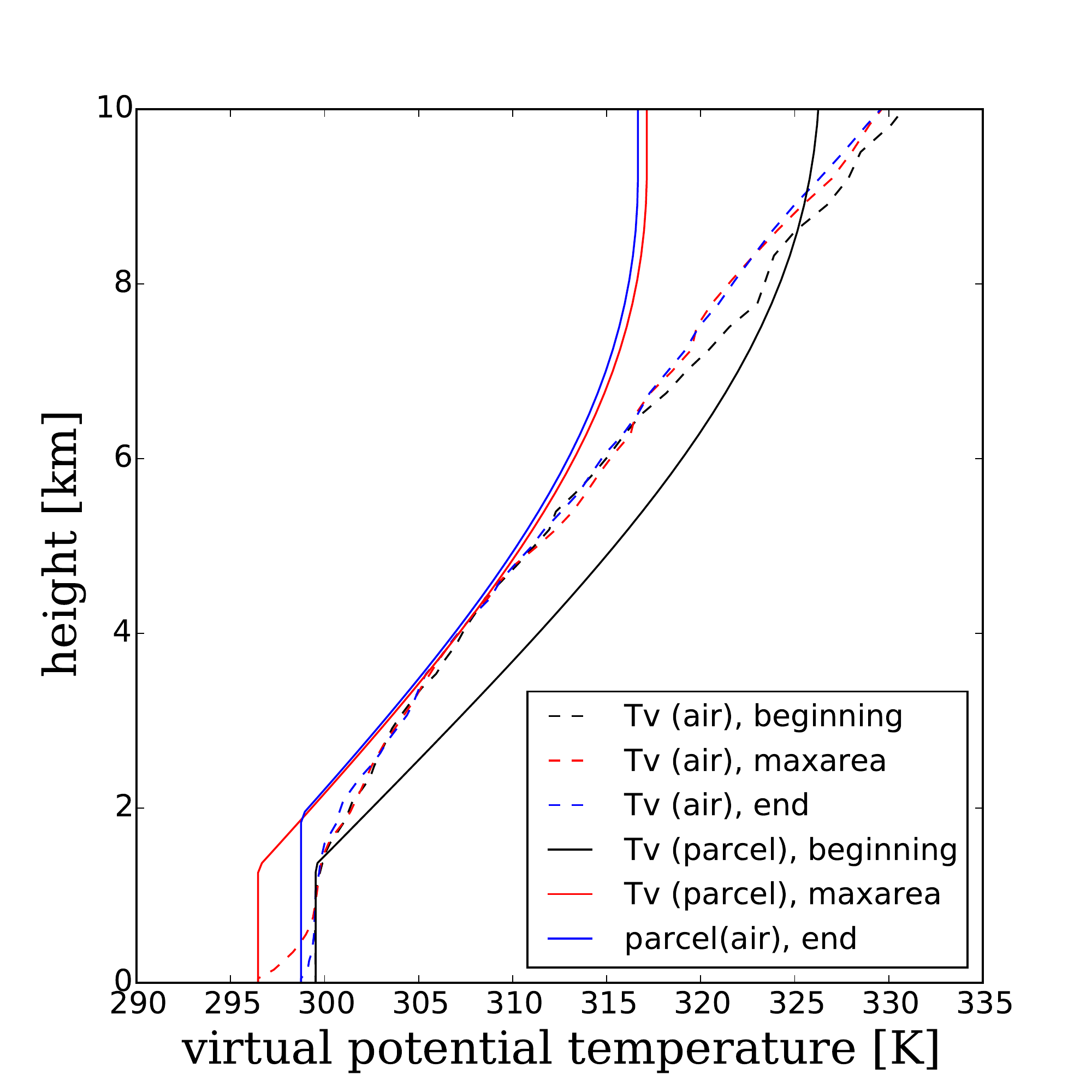}
    \includegraphics[width=6.0cm,trim={0 0 0 0} ,clip]{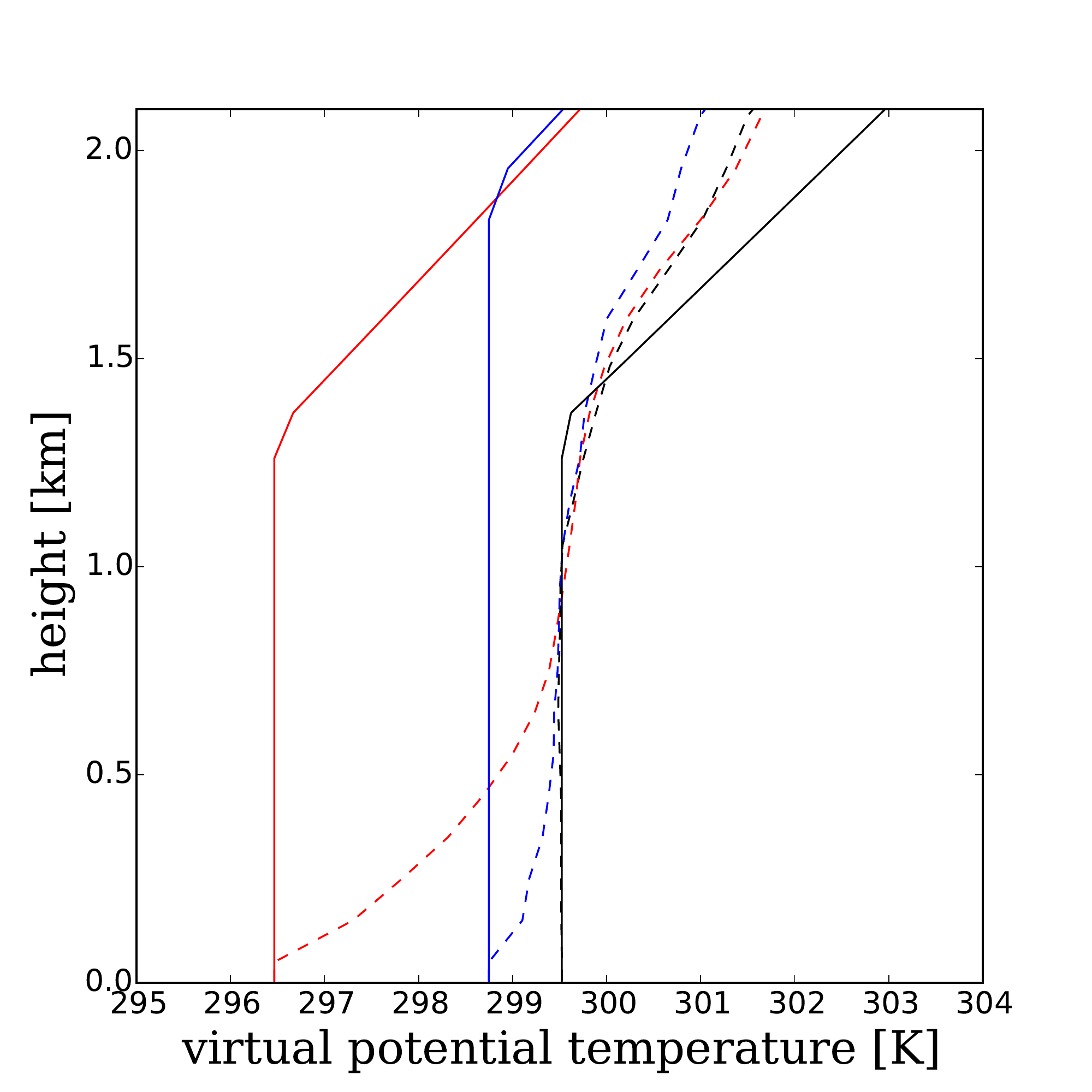}
    \caption{{\bf Vertical profiles of virtual potential temperature $\theta_v$ for a selected solitary track.}
    Dashed curves show the profile of $\theta_{v}$ at the column of the center of mass of the event, while solid curves show the profile of an idealized air parcel that follows a pseudo adiabat $\theta_{v,palcel}$, at the beginning of the event (left), after 30 min when the track reaches its maximum extent (center), and after 70 minutes at the time when surface rain ends (right). The area between both curves where $\theta_{v,parcel}>\theta_{v}$ represents CAPE, while the area close to the surface, where $\theta_{v,parcel}<\theta_{v}$, represents CIN. Left panel: Profiles from surface up to a height of 10 km; right panel: Zoom into boundary layer up to a height of 2 km. While at the beginning of the events CAPE is large and CIN is small, after 30 min CAPE is large, mainly because of the drop in surface temperature due to the cold pool, while CAPE completely vanishes (note that this is the case only for a part of the tracks). At the end of the track, the cold pool has already weakened and a drop in CIN is visible. Note also that the change in atmospheric profile above the boundary layer (i.e. the dashed lines) is relatively small and therefore contributes only weakly to the change in CAPE. The increase of the level of free convection at the end of the tracks indicated a drying of the boundary layer.
    }
    \label{fig:vpt_profiles}
\end{figure}

\begin{figure}
    \centering
    \includegraphics[width=3.6cm,trim={0 2.35cm 0  0},clip]{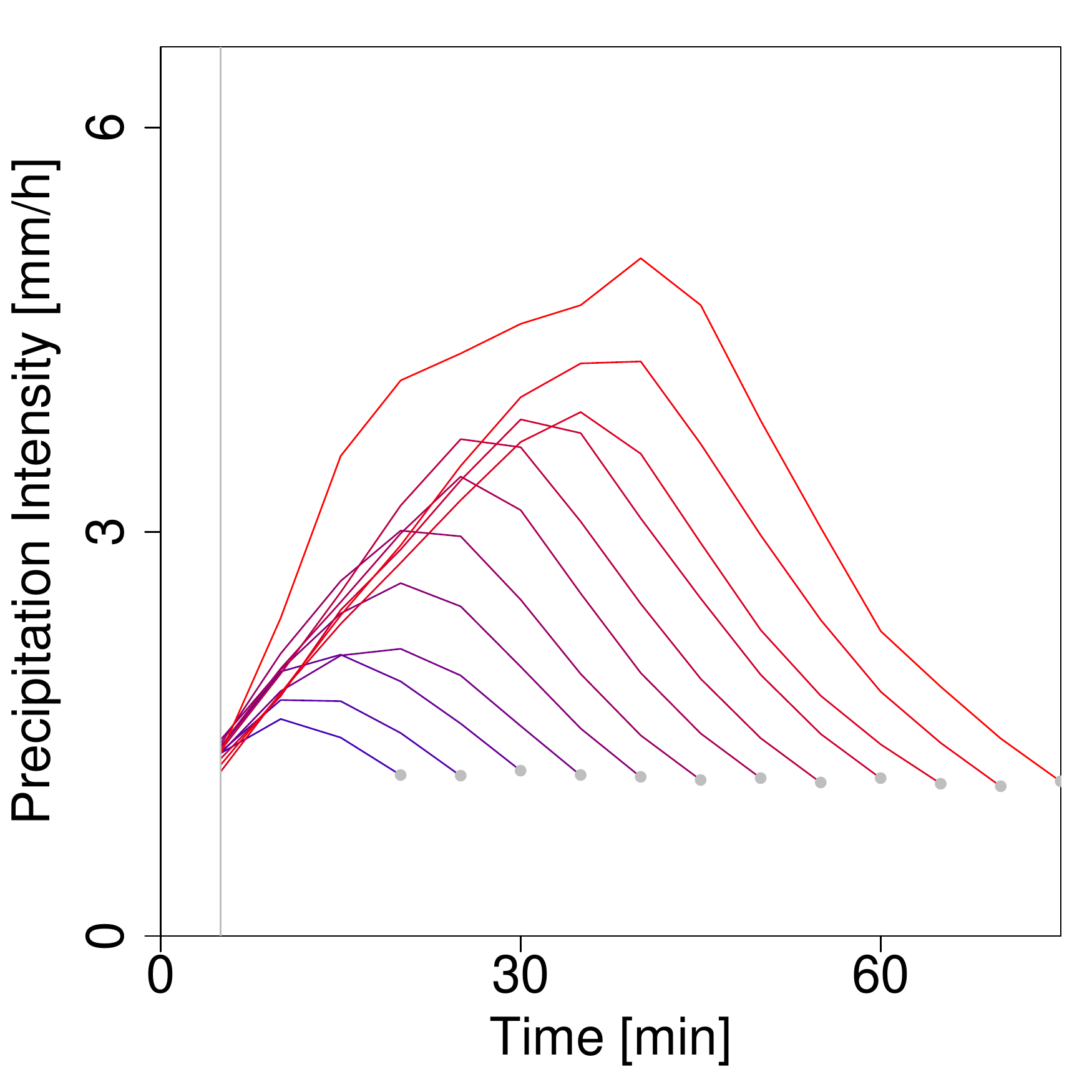}
    \includegraphics[width=3.2cm,trim={2.1cm 2.35cm 0  0},clip]{{cond_solitary-solitary_p2K_1.0_precip_vs_time}.pdf}
    \includegraphics[width=3.2cm,trim={2.1cm 2.35cm 0  0},clip]{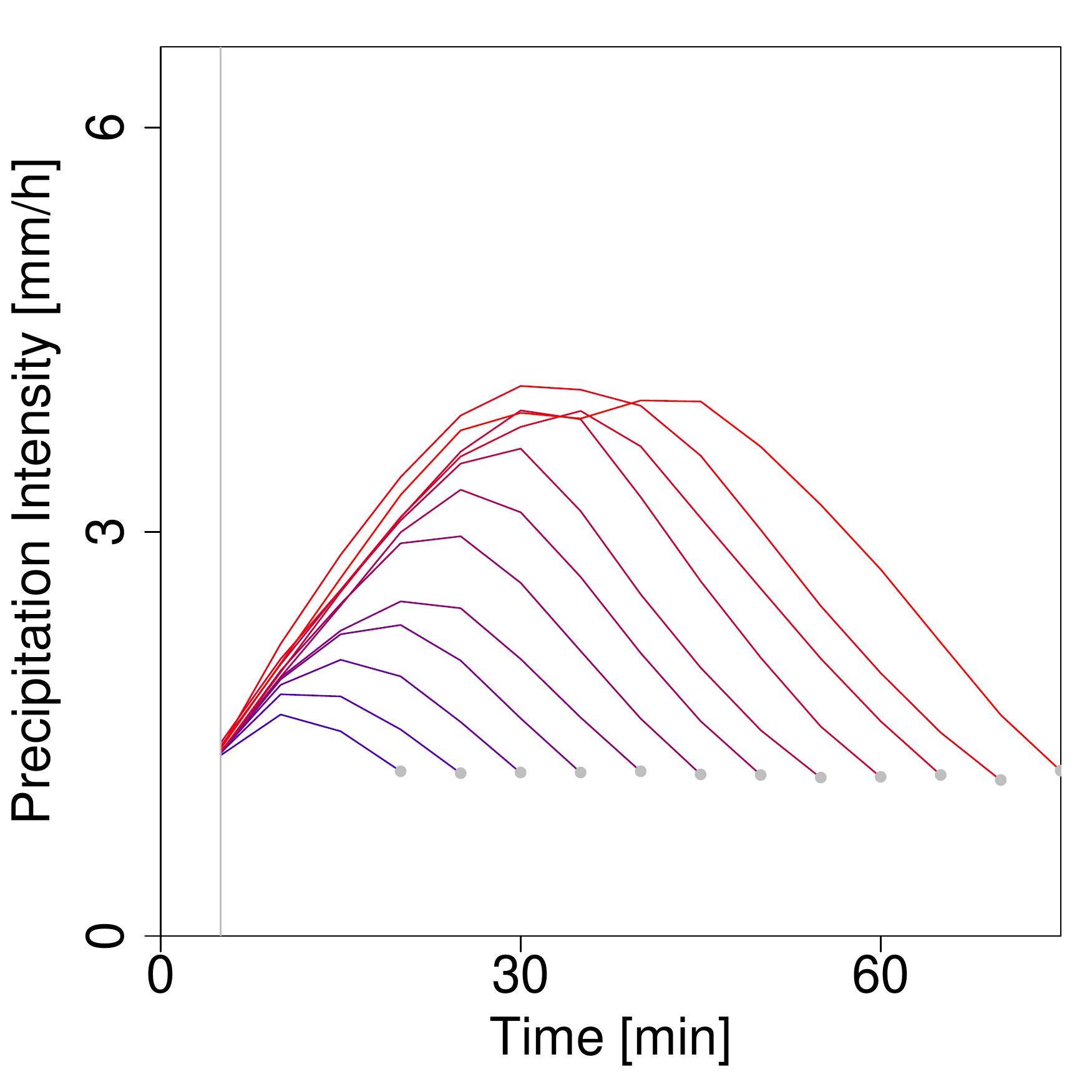}
    \includegraphics[width=3.2cm,trim={2.1cm 2.35cm 0  0},clip]{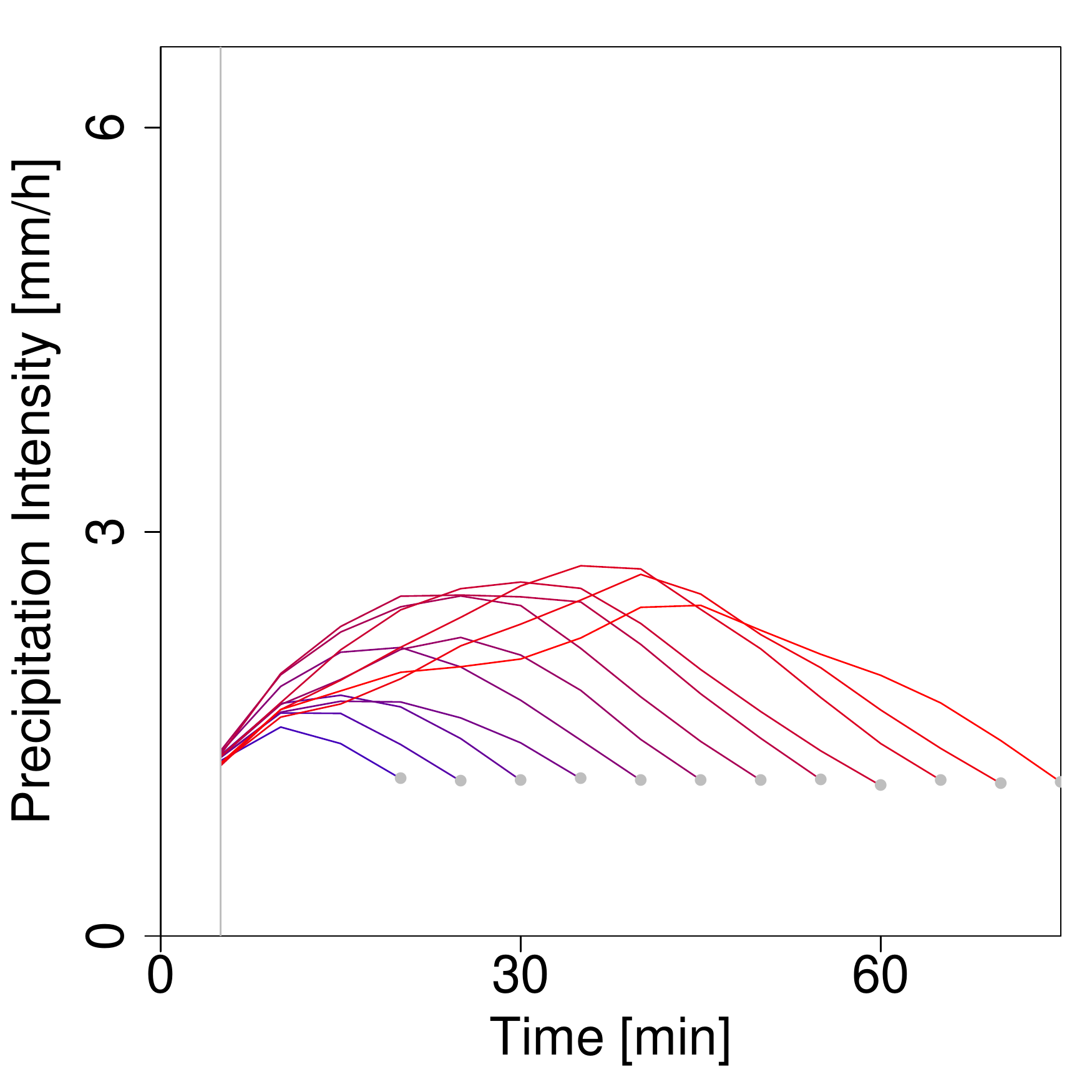}
    \\
    \vspace{-0.0cm}
    \includegraphics[width=3.6cm,trim={0 2.35cm 0  0},clip]{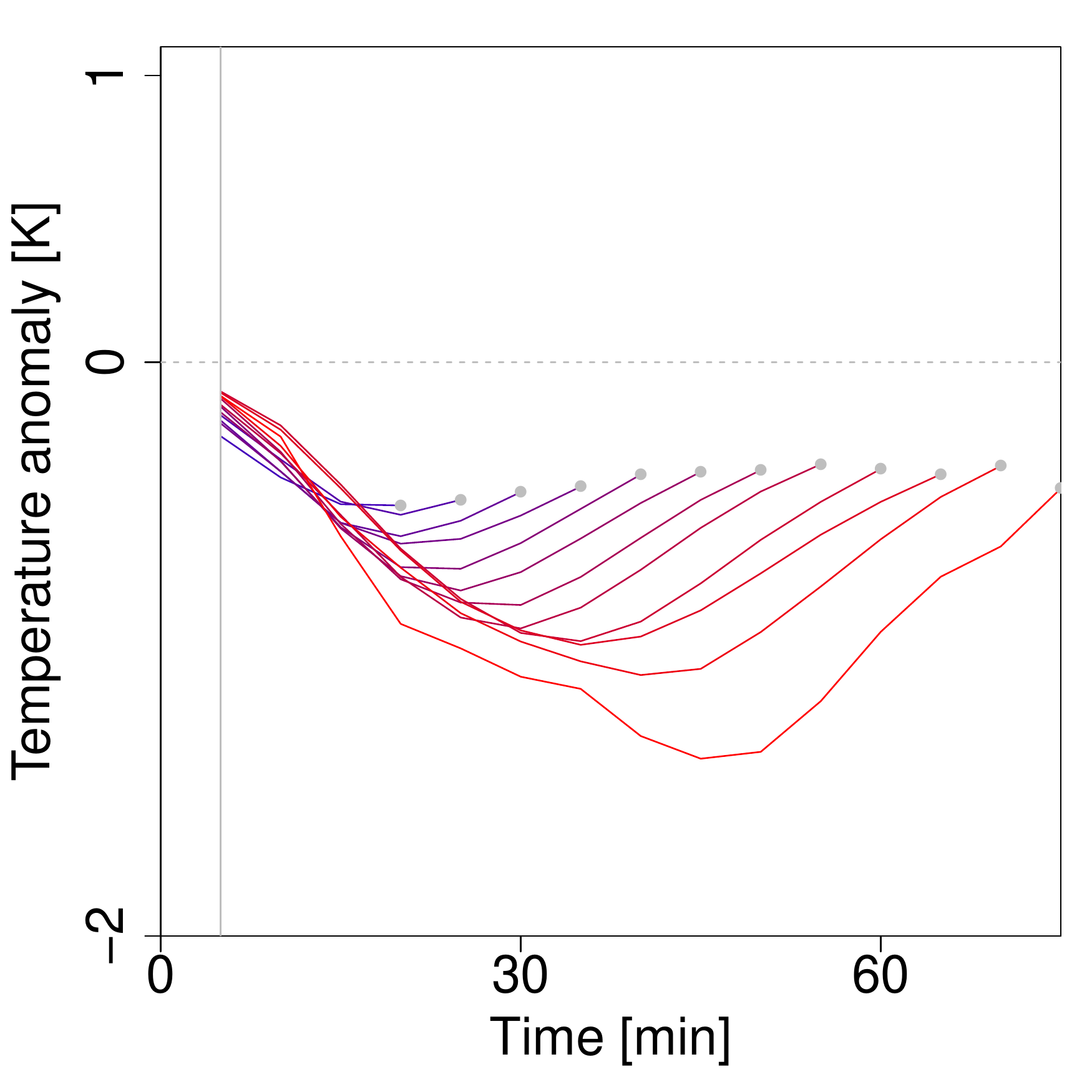}
    \includegraphics[width=3.2cm,trim={2.1cm 2.35cm 0  0},clip]{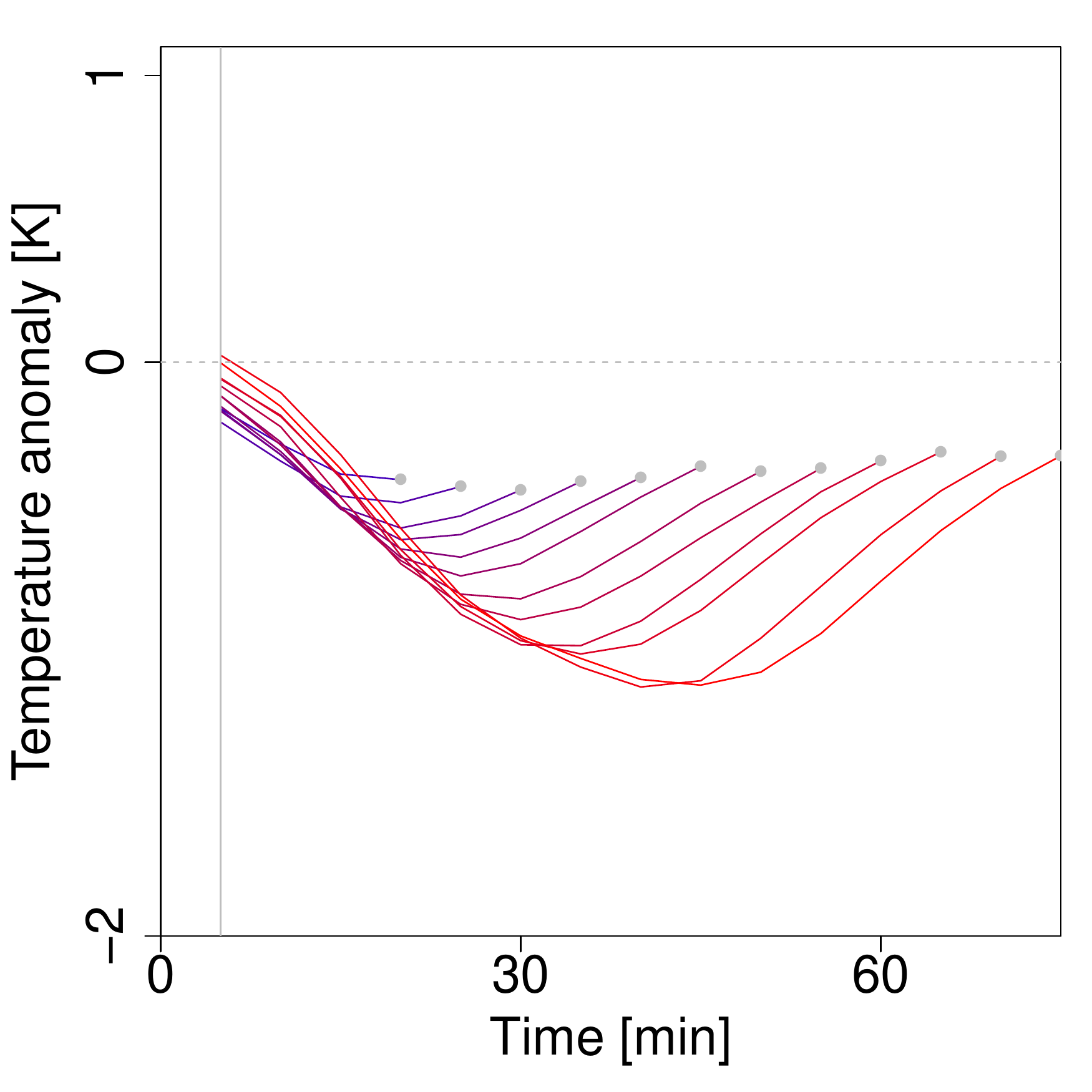}
    \includegraphics[width=3.2cm,trim={2.1cm 2.35cm 0  0},clip]{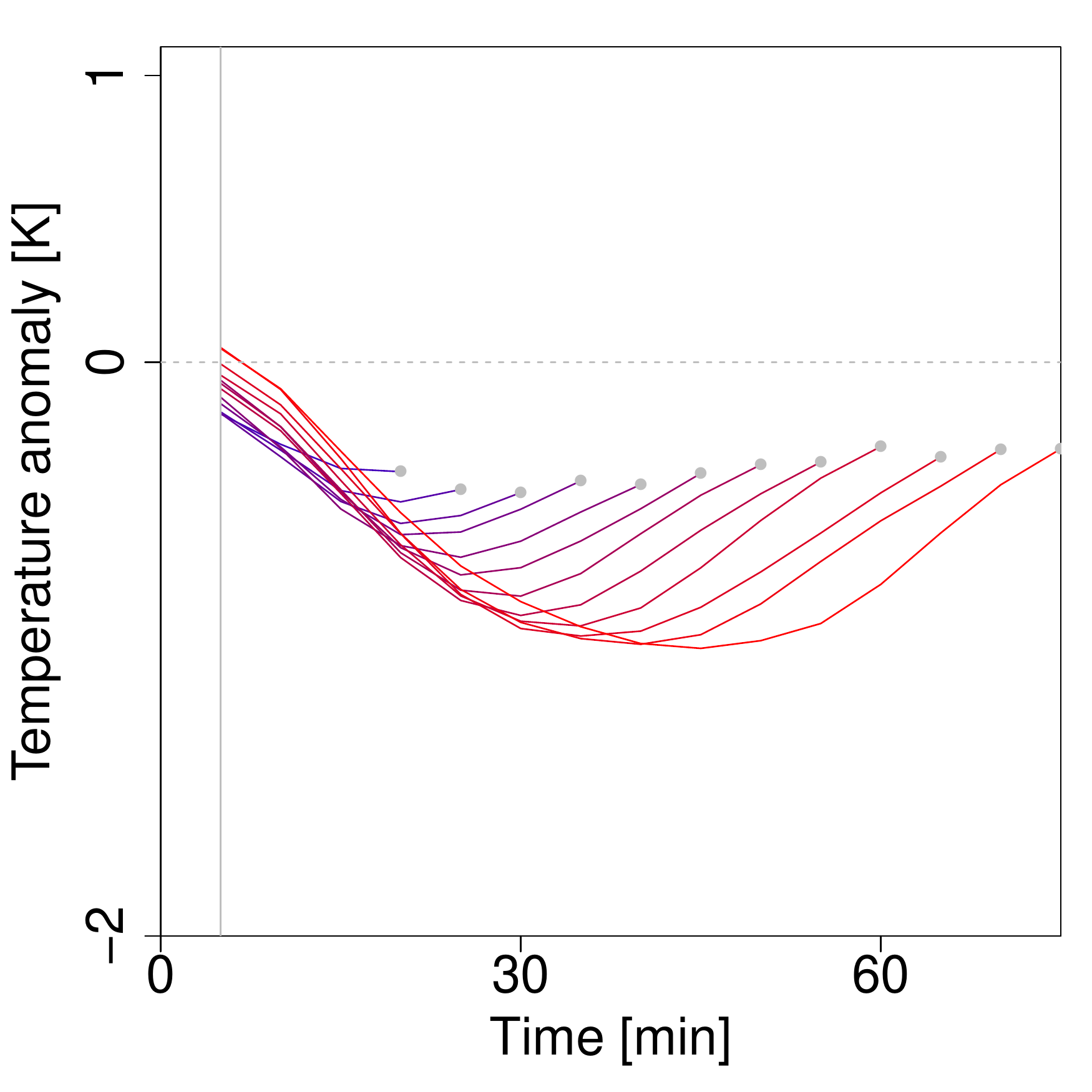}
    \includegraphics[width=3.2cm,trim={2.1cm 2.35cm 0  0},clip]{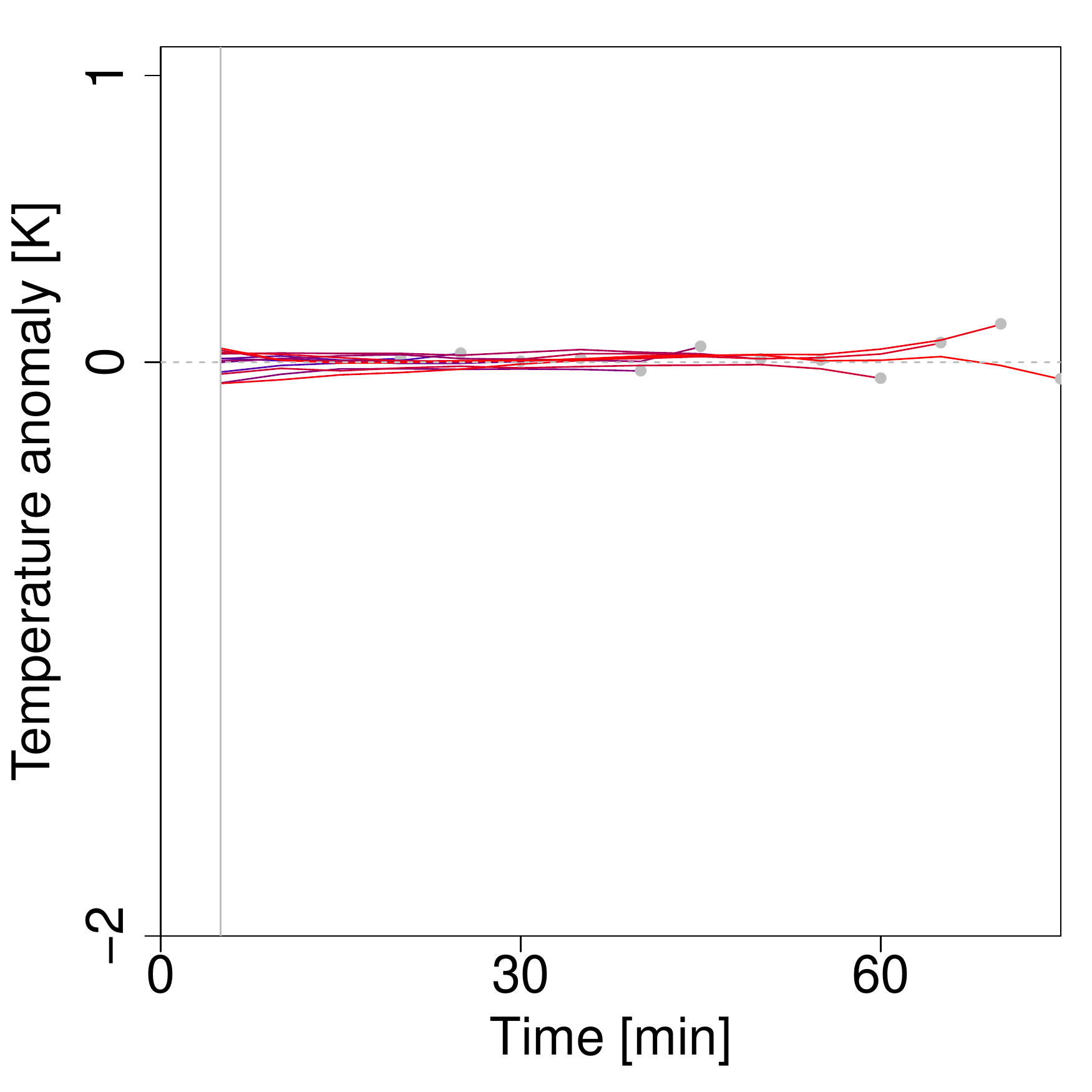}
    \\
    \vspace{-0.0cm}
    \includegraphics[width=3.6cm,trim={0 2.35cm 0  0},clip]{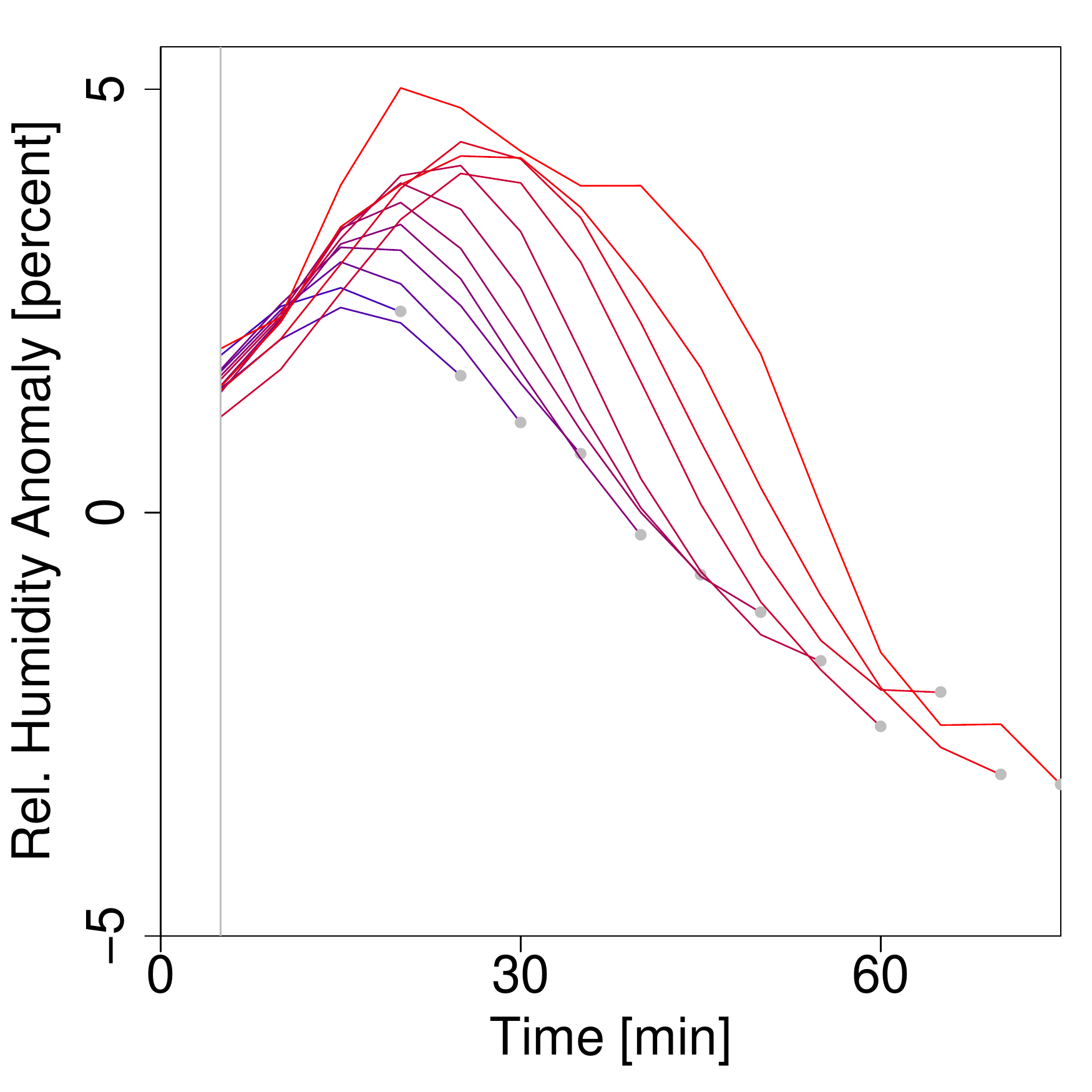}
    \includegraphics[width=3.2cm,trim={2.1cm 2.35cm 0  0},clip]{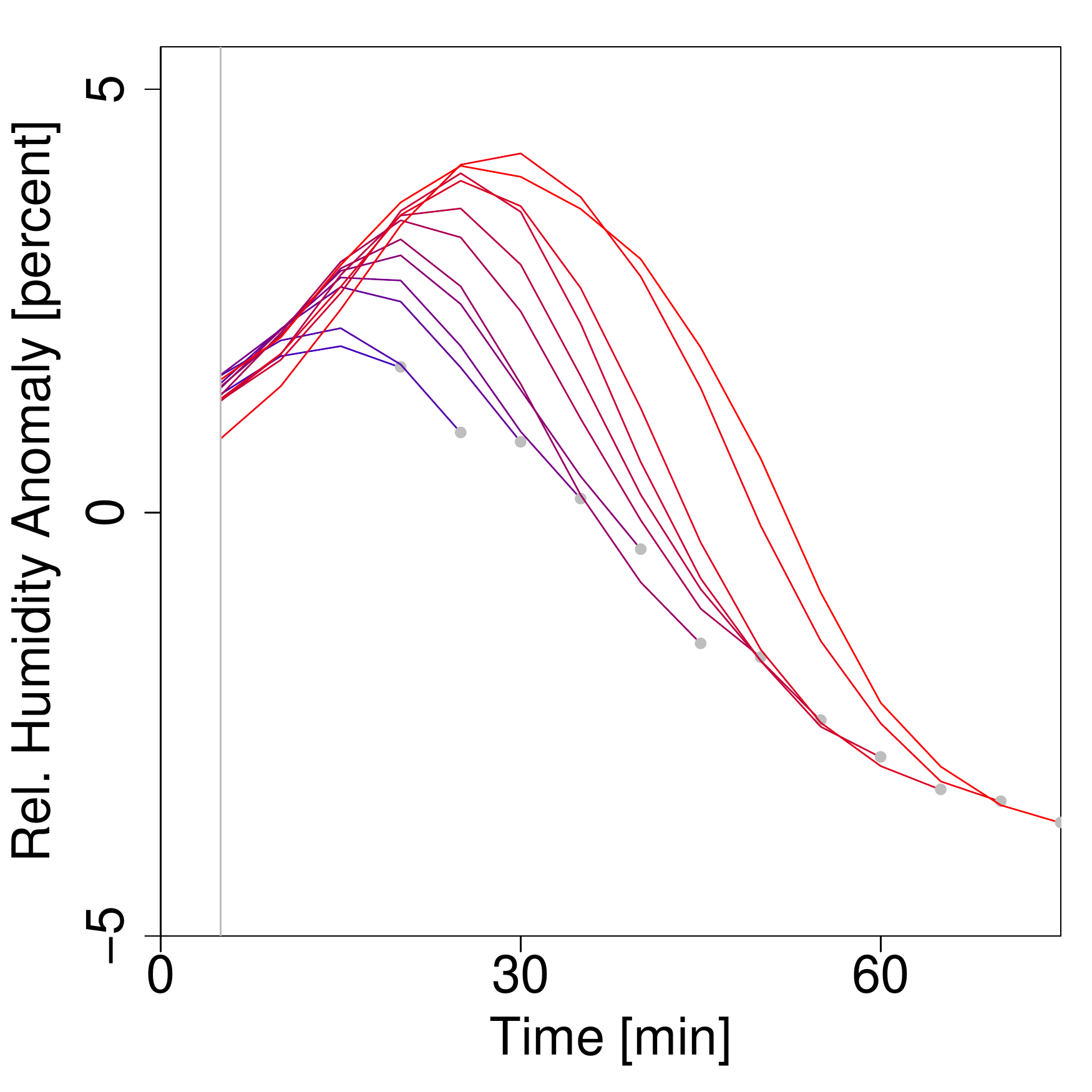}
    \includegraphics[width=3.2cm,trim={2.1cm 2.35cm 0  0},clip]{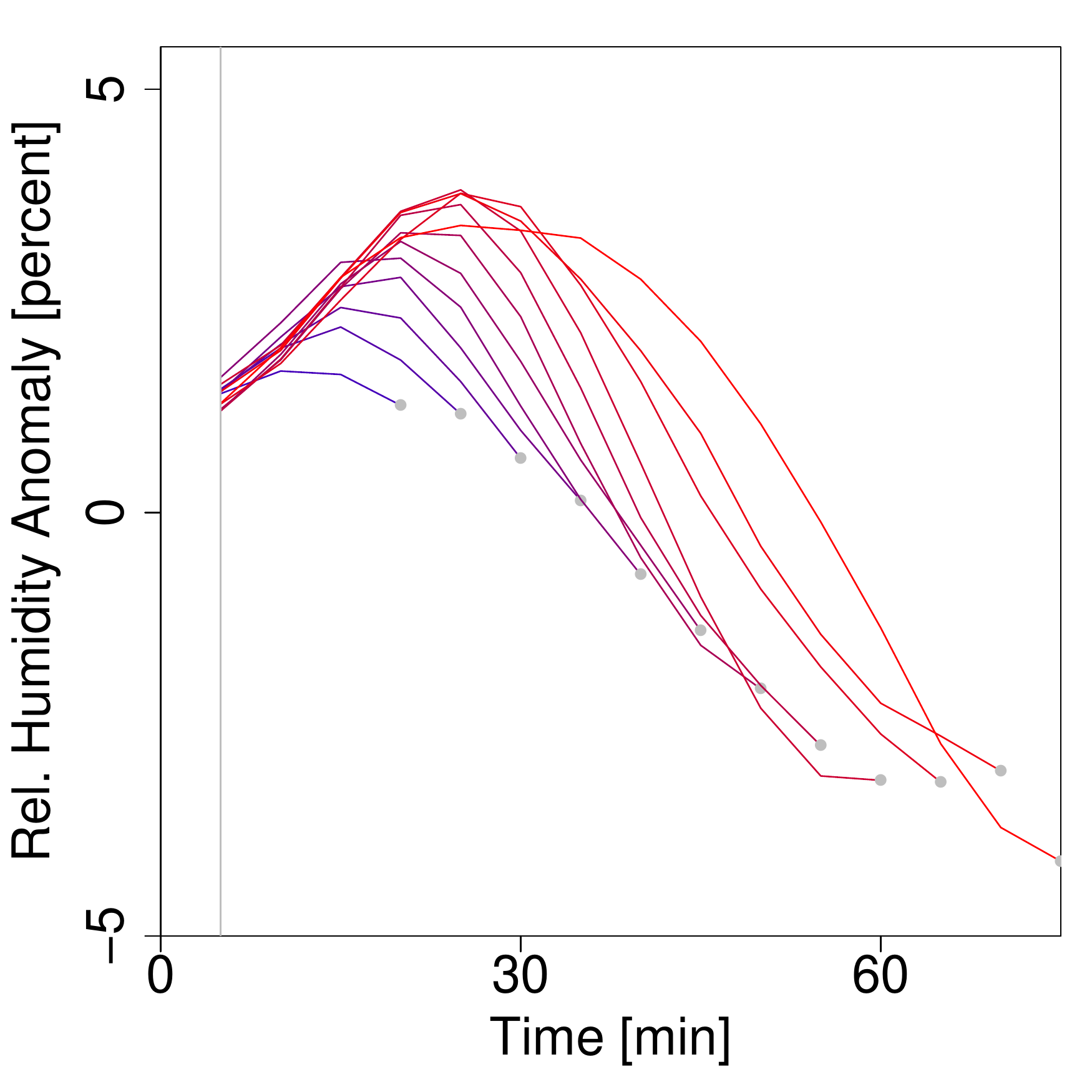}
    \includegraphics[width=3.2cm,trim={2.1cm 2.35cm 0  0},clip]{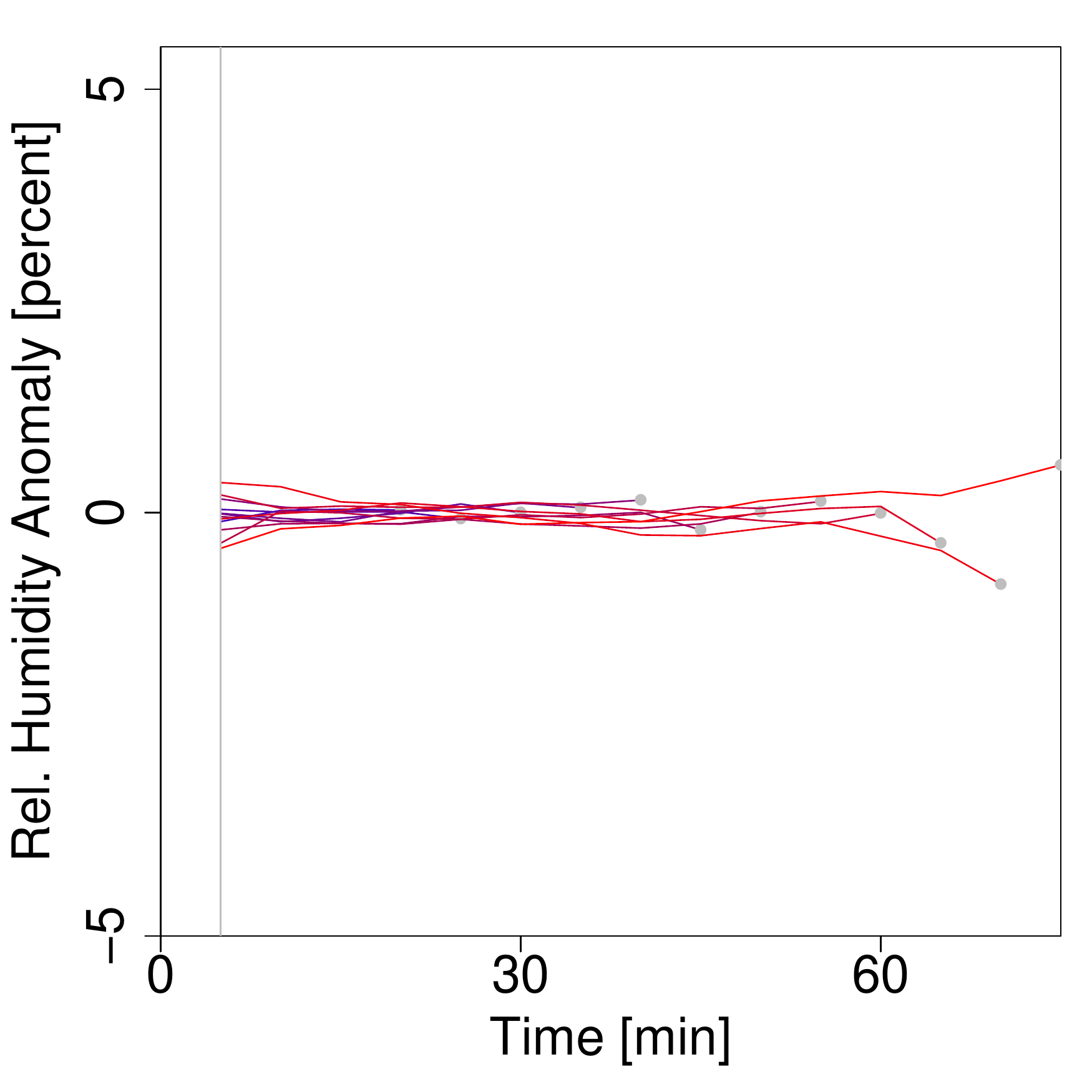}
    \caption{{\bf Auxiliary field time dependence.}
    Integsity (top row), near-surface temperature (center row), and relative humidity (bottom row) as function of time. Columns from left to right show solitary tracks for CTR, P2K, P4K, and OMEGA, each for $\theta=1.0$.
    }
    \label{fig:time_dependence_aux}
\end{figure}

\begin{figure}[b]
    \centering
    \includegraphics[width=10cm,clip]{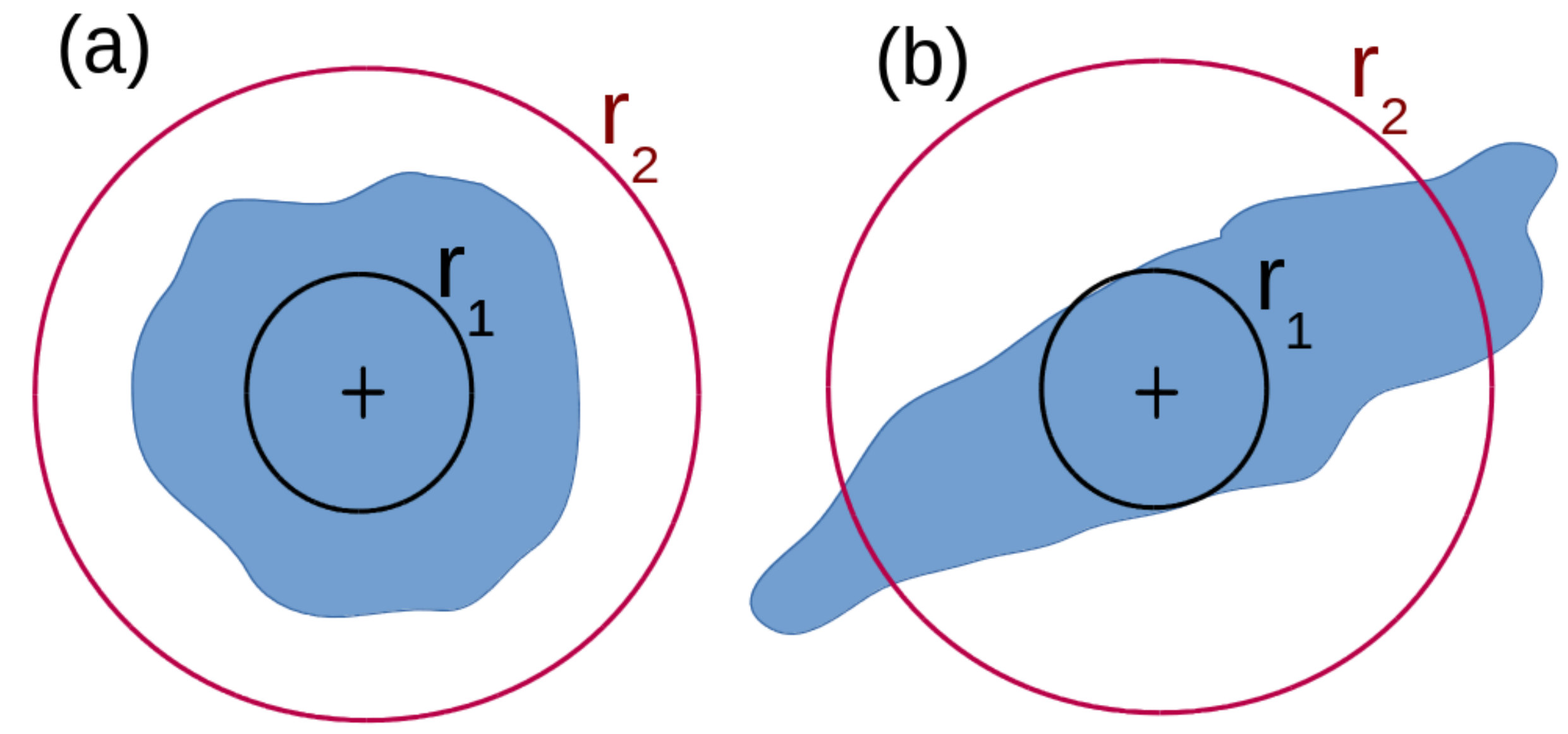}
    \caption{Sketch to illustrate the tails of the radial profiles seen in Fig.~\ref{fig:radial_profiles}. Consider two precipitation objects of comparable size, one which has nearly circular shape (a) and one that strongly deviates from a circular shape (b). The black circles with small radius $r_1$ around both cell centers lie completely inside the rain areas, so the mean intensity at this radius will be averaged around the entire circle. However, the red circle with large radius $r_2$ is completely outside the object (a), such that the intensity there will be zero, while in (b) the red circle partly intersects with the objects. Outside this intersection the intensity is by definition zero, such that an averaging along this circle will lead to a small mean intensity and thus contribute to the tail. Averaging over many circular shaped objects of type (a) and few longer objects of comparable size of type (b) leads to the tail of the profiles, especially for larger areas.}
    \label{fig:conemodel_sketch}
\end{figure}


\end{document}